\documentclass[twocolumn]{aastex631}

\usepackage{graphicx}	
\usepackage{amsmath}	
\usepackage{amssymb}	

\shorttitle{Outbursts and chondrites}
\shortauthors{Li et al.}

\graphicspath{{./}{figures/}}

\begin{document}

\title{Stellar outbursts and chondrite composition}

\author{Min Li}
\affiliation{College of Physics, Jilin Normal University, Siping, Jilin 136000, China}
\affiliation{Department of Physics and Astronomy, University of Nevada, Las Vegas, 4505 S. Maryland Pkwy, Las Vegas, NV 89154, USA}
\affiliation{Key Laboratory of Functional Materials Physics and Chemistry of the Ministry of Education, Jilin Normal University, Changchun, Jilin 130103, China}

\author{Zhaohuan Zhu}
\affiliation{Department of Physics and Astronomy, University of Nevada, Las Vegas, 4505 S. Maryland Pkwy, Las Vegas, NV 89154, USA}
\affiliation{Nevada Center for Astrophysics, University of Nevada, Las Vegas, 4505 S. Maryland Pkwy, Las Vegas, NV 89154, USA}

\author{Shichun Huang}
\affiliation{Department of Earth and Planetary Sciences, University of Tennessee, Knoxville, 1621 Cumberland Ave, Knoxville, TN 37996, USA}

\author{Ning Sui}
\affiliation{College of Physics, Jilin University, Changchun, Jilin 130012, China}

\author{Michail I. Petaev}
\affiliation{Department of Earth and Planetary Sciences, Harvard University, 20 Oxford Street, Cambridge, MA 02138, USA}

\author{Jason H. Steffen}
\affiliation{Department of Physics and Astronomy, University of Nevada, Las Vegas, 4505 S. Maryland Pkwy, Las Vegas, NV 89154, USA}
\affiliation{Nevada Center for Astrophysics, University of Nevada, Las Vegas, 4505 S. Maryland Pkwy, Las Vegas, NV 89154, USA}
\correspondingauthor{Jason H. Steffen}
\email{jason.steffen@unlv.edu}

\begin{abstract}

The temperatures of observed protoplanetary disks are not sufficiently high to produce the accretion rate needed to form stars, nor are they sufficient to explain the volatile depletion patterns in CM, CO, and CV chondrites and terrestrial planets.  We revisit the role that stellar outbursts, caused by high accretion episodes, play in resolving these two issues.  These outbursts provide the necessary mass to form the star during the disk lifetime, and provide enough heat to vaporize planet-forming materials.  We show that these outbursts can reproduce the observed chondrite abundances at distances near one AU.  These outbursts would also affect the growth of calcium-aluminum-rich inclusions (CAIs) and the isotopic compositions of carbonaceous and non-carbonaceous chondrites.

\end{abstract}


\section{Introduction} \label{sec:intro}

The formation and evolution of a star and its associated protoplanetary disk are not fully understood.  For example, the ``Luminosity Problem'' is a long-standing issue in our understanding of the formation of low-mass stars like the Sun.  The observed protostellar luminosities are nearly an order-of-magnitude lower than the luminosity expected from steady-state accretion of material from the protostellar disk.  These observations imply an accretion rate that is too low for stars to form within the disk lifetimes.  The Luminosity Problem is an issue first discovered by \cite{Kenyon:1990} but drew great attention in the star formation community when it was confirmed by the Spitzer C2D survey \citep{Evans:2009}.

As another example, the chemical compositions and mineralogy of chondrites in the Solar System require a hot Solar nebula---of order 2,000 K---at and beyond 1 AU \citep{Cassen:1996,Ciesla:2008,Li:2020}.  Starting from this hot environment, the cooling and evolving disk produces chondrites (and planets) in the Solar System that are depleted in volatile and enriched in refractory elements \citep{Cassen:1996,Bond:2010,Elser:2012,Pignatale:2016,Li:2020}.  However, such high disk temperatures, with their implied high disk luminosities, have not been observed around other young stellar objects, as in the ``Luminosity Problem''.  Moreover, from the theory side, the midplane temperatures of protoplanetary disks, when modeled from the collapse of molecular cloud cores, rarely reach such high temperatures at 1 AU \citep{Li:2021}.

Since star formation and planetesimal formation occur simultaneously, whatever solution is found for the stellar ``Luminosity Problem'' must also affect the ongoing physiochemical processes in the protoplanetary disk that form the planetesimals.  One solution is that stars may undergo multiple episodes of high disk accretion rates \citep{Kenyon:1990}.  The stars in this scenario spend most of their time accreting slowly, with low luminosities, but undergo bursts of rapid disk accretion where most of the stellar mass is gained.  Although some work implies that the ``Luminosity Problem'' may not be as serious as once inferred \citep{Li:2018}, several large outbursts are still needed in order to match the latest theoretical models to observed disks.

We indeed observe many active, young stellar objects exhibiting such outbursts.  They range from relatively mild ones (Exors outbursts), to large outbursts that brighten the system by five magnitudes in the V-band (FUors outbursts) \citep{Hartmann:1996,Audard:2014}.  Several ongoing surveys will constrain the frequency of FUors \citep{Hillenbrand:2015,Connelley:2018,Fischer:2023}.  The existence of these outbursts may resolve the Luminosity Problem, but challenges the traditional, steady-state disk accretion picture.  If common, these outbursts may significantly alter our understanding of planet formation \citep{DV2012}, binary formation \citep{Stamatellos2012}, the luminosity distribution in young clusters \citep{Baraffe2009}, and disk chemistry \citep{VB2012,Forbes:2021}.

FUors outbursts release sufficient energy to evaporate even the most refractory elements in the disk midplane.  It is observed that, during an outburst episode, the inner disk within 1 AU is extremely hot---with surface temperatures well above 2000 K \citep{Zhu:2007,Bae2014, Zhu:2020}.  Given this information, these outbursts may provide the necessary conditions to produce the observed chemical trends in Solar System chondrites and planets.  In this paper, we couple a disk model that produces FUors outbursts with a chemical condensation model to examine how this proposed solution to the Luminosity Problem may also resolve the tension between the necessary temperature of the protoplanetary disk inferred from chondrites, and the temperatures seen in observations and theoretical modeling.

Although there are some uncertainties regarding the triggering mechanism for FUors outbursts (with proposed mechanisms that include binary interactions \citep{Bonnell:1992,Clarke:1996}, clump accretion, \citep{Vorobyov:2005} and disk instability \citep{Armitage:2001}), the disk instability model is well studied and produces quantitative predictions that are consistent with observational constraints on the size of the hot disk (AU scale) \citep{Zhu2010b}. For this work, we adopt this disk instability approach, which can generate a high-temperature inner disk that is capable of affecting the disk chemistry.

Our model is built on first-principle disk accretion mechanisms, and begins at the collapse of the molecular cloud core \citep{Zhu2010b,Li:2015}.  Here, two different instabilities combine to produce episodic bursts of accretion \citep{Zhu:2009,Martin:2011,Bae2013}.  The gravitational instability (GI, \cite{Durisen:2007}) operates in the outer disk when the disk is massive.  It funnels material from the outer disk to the inner disk where it accumulates.  Viscous heating raises the temperature of the material until it is high enough for thermal ionization to trigger magnetorotational instability (MRI, \cite{Balbus:1998}).  At that point, the coupling of the magnetic field to the ionized gas drives angular momentum outward and the orbiting material falls into the central star.  This rapid accretion of material leads to the outburst.

We combine this disk model with a dust condensation model, GRAINS, which calculates the equilibrium partitioning of 33 chemical elements between the gaseous and condensed phases for given temperature and pressure conditions \citep{Petaev:2009}.  Over time, some material condenses and decouples from the disk while more volatile material remains in the gaseous state and advects with the evolving disk.  In this manner, we self-consistently calculate the hydrodynamical and chemical evolution of the protoplanetary disk.

Throughout this work, we compare the results of our model with a standard $\alpha$-disk model that has a constant, uniform viscosity and does not produce outbursts.  We find that our outburst model yields good agreement between the modeled compositions of the condensed planetesimal material in the midplane of the disk and the observed volatile depletion patterns in CM, CO, and CV chondrites, and terrestrial planets.  The constant-$\alpha$ model does not produce similar agreement.

In Section \ref{sec:disk}, we describe the methods we use to model our Solar System's protoplanetary disk and the disk outbursts.  In Section \ref{sec:results}, we compare the evolution of the disk temperature and surface density for the outburst model with the constant $\alpha$-disk model.  We then examine the effects that the outbursts have on the chemical evolution of the disk and compare those with measured chemical compositions of chondrites.  Finally in Section \ref{sec:discussion}, we discuss some implications that this model may have for the thermal history of the Solar System, the formation of chondrites and calcium-aluminum-rich inclusions (CAIs), and the radial evolution of the ice and rock lines outside of which rocky and icy planetesimals form.

\section{Disk evolution model}\label{sec:disk}

We use the standard viscous disk theory to study the evolution of the protoplanetary disk that forms from the collapse of a molecular cloud core (MCC) \cite{Li:2015}.  The evolution of the disk surface density is given by:
\begin{eqnarray}\label{equ.diff}
\frac{\partial \Sigma(R,t)}{\partial t}
&=&\frac{3}{R} \frac{\partial}{\partial R} \left[ R^{1/2} \frac{\partial}{\partial R} (\Sigma \nu R^{1/2}) \right] +S(R,t)\nonumber \\
&+&S(R,t)\left\{2-3\left[\frac{R}{R_{d}(t)}\right]^{1/2}+\frac{R/R_{d}(t)}{1+[R/R_{d}(t)]^{1/2}}\right\}.
\end{eqnarray}
Here $\Sigma(R,t)$ is the gas surface density of the disk at radius $R$ and time $t$, and $\nu$ is the kinematic viscosity.  The third term on the right hand side of Equation (\ref{equ.diff}) arises from the difference between the specific angular momentum of the infalling material and that of the material in the disk. $S(R,t)$ is the mass influx onto the disk and protostar system  \citep{Nakamoto:1994}:
\begin{eqnarray}\label{equ.inf}
S\left(R,t\right) =\left\{
\begin{aligned}
& \frac{\dot{M_{\text{in}}}}{4\pi R R_{\mathrm{d}}\left(t\right)}
\left[1-\frac{R}{R_{\mathrm{d}}\left(t\right)}\right]^{-1/2} & {\rm if\ }
\frac{R}{R_{\mathrm{d}}\left(t\right)}<1  \\
 &  0\ & {\rm otherwise}
\end{aligned}
\right.
\end{eqnarray}
where $\dot{M_{in}}$ is the mass infall rate of an isothermal sphere with a temperature of $T_{\rm C}$ \citep{Shu:1977}
\begin{equation}\label{equ.mdot}
\dot{M_{in}} = \frac{0.975}{G}\left(\frac{\mathcal{R}}{\mu}\right)^{3/2}T_{\rm C}^{3/2},
\end{equation}
where $G$ is the gravitational constant, $\mathcal{R}$ is the gas constant, and $\mu = 2.33$ is the mean molecular mass.  $R_{\mathrm{d}}$ is the centrifugal radius of the infalling material.  This radius increases with time as the higher angular momentum material in the MCC falls to the disk later in its evolution,
\begin{equation}\label{equ.rd}
R_{\mathrm{d}}(t) = 31
\left(\frac{\omega_{\rm C}}{10^{-14} {\rm\ s} ^{-1} } \right)^{2}
\left(\frac{T_{\rm C}}{10 {\rm\ K}} \right)^{1/2}
\left(\frac{t}{5\times10^{5} {\rm\ yr}} \right)^{3} {\rm AU},
\end{equation}
where $\omega_{\rm C}$ is the angular velocity of the MCC.  The values we use in this equation come from observations of MCCs.  Specifically, the rotation speeds of MCCs are generally a few times $10^{-14}$ s$^{-1}$ \citep{Jijina:1999}, their temperatures are typically a few tens of Kelvins \citep{Goodman:1993}, and their lifetimes are generally a few times $10^{5}$ years \citep{Li:2016,Strom:1989}.

We use the $\alpha$-prescription \citep{Shakura:1973} to calculate the viscosity, $\nu=\alpha c_s H$, where $\alpha$ is a dimensionless parameter less than 1, $H$ is the half thickness of the gas disk, $c_s=\sqrt{\mathcal{R} T/\mu}$ is the sound speed, and $T$ is the temperature of the mid-plane of the disk. To calculate the midplane temperature of the disk, we adopt a similar equation used in \citet{Cannizzo:1993} and \citet{Armitage:2001}
\begin{equation}\label{equ.tm}
\frac{\partial T_c}{\partial t}=\frac{2\left(Q_+-Q_-\right)}{c_{\rm p} \Sigma}.
\end{equation}
Here $c_{\rm p}$ is the specific heat. cyanThe heating sources include viscous heating
\begin{equation}\label{equ.Qplus}
Q_{\rm vis}=\frac{9}{8}\nu\Sigma\Omega^2,
\end{equation}
and heating by infalling material during the cloud core collapse, $Q_{\rm infall}$. Irradiation from the central star, disk accretion, and envelope environment are also considered. Detailed calculation method is similar to that in \citet{Bae2013}.
and $Q_-$ is the local cooling rate
\begin{equation}\label{equ.Qminus}
Q_-=\sigma T_{\rm e}^{4},
\end{equation}
where $\sigma$ is the Stefan--Boltzmann constant, and $T_{\rm e}$ is the local temperature of the disk if treated as a blackbody. The midplane temperature from Equation \ref{equ.tm} is related to the effective temperature from Equation \ref{equ.Qminus} through the disk surface density and opacity. We use the same method as in \citet{Armitage:2001} to calculate the opacity, which comes from \citet{BellLin1994} for high temperatures and \citet{Bell:1997} for low temperatures.

The key parameter governing the disk evolution is $\alpha$.  As a zeroth-order approximation, $\alpha$ is assumed to be constant throughout the disk \citep{Shakura:1973}.  In \cite{Li:2021}, we studied dust condensation using the constant-$\alpha$ model and showed that the model does not heat the disk enough to vaporize the moderately volatile elements---implying that they are not fractionated from the refractory elements.  Consequently, the constant-$\alpha$ model does not yield elemental patterns consistent with those observed in chondrites.  We consider refractory elements to be those with 50\% condensation temperatures higher than Silicon.  Similarly, we define an element as moderately volatile when the 50\% condensation temperature is between 1100 and $\sim$ 1300 K.

We know the constant-$\alpha$ model is over-simplified.  Every disk instability that can lead to angular momentum transport is initiated by its own set of conditions and causes its own change in the local value of $\alpha$.  The mismatch of the disk accretion rates by the different instability mechanisms can lead to accretion outbursts, as described in the introduction.  To account for this in our outburst model, we specify different, effective $\alpha$ values that apply under different disk conditions.  For MRI to operate, the disk needs to be ionized, either by thermal or non-thermal (e.g. cosmic ray and X-ray) ionization.  Following the model by \cite{gammie1996}, we assume that the critical temperature for this thermal ionization is 1500 K.  MRI can also operate when the disk's surface density is less than 200 g cm$^{-2}$ due to non-thermal ionization.

When MRI is active, the $\alpha$ parameter is set to 0.001, which is the same as that in the fiducial constant $\alpha$ model in \cite{Li:2021}, allowing us a proper comparison with the constant $\alpha$ model. Although this value is smaller than those used in previous outburst models, it can still lead to the correct outburst amplitudes with some adjustment to the ionization temperature \citep{Zhu2010a}.  When MRI is not active, we assume that $\alpha$ can drop down to a minimum value of $4\times10^{-5}$ in the MRI ``deadzone''.  This non-zero floor can be caused by other hydrodynamical instabilities in the disk (e.g. streaming instability \citep{Johansen2007} or vertical shear instability \citep{Nelson2013}).  The parameters we use in our default models include: the critical temperature for the thermal ionization is 1500 K and $\alpha=4\times10^{-5}$ in the dead zone (default outburst model in the future). We also show the results for a ``high ionization temperature model'' where the critical temperature for the thermal ionization is 1800 K with $\alpha=3\times10^{-5}$ in the dead zone.

For Gravitational Instability (GI) to operate, the disk needs to be massive.  We adopt the formulae in \cite{Armitage:2001} for GI with $\alpha_{GI}$=0.001 and $Q_{crit}$=2.  Thus, GI only leads to angular momentum transport when the Toomre $Q<2$.  Although outbursts can occur for a wide range of parameter values, the choice of these values affects the outburst strength and frequency \citep{Zhu2010a}. Thus, we choose values that lead to outbursts that match observed FU Orionis systems.

We start the GRAINS chemistry calculation during the last outburst at the moment where the temperature at 1 AU is at its highest value.  We then follow the chemistry throughout the simulation to 1 Myrs.  Here, we assume that the outburst essentially resets the chemistry. An investigation of the cumulative effect of multiple outbursts on the condensed material is left for future work.  A detailed description of our algorithm tracking the chemical evolution of the disk can be found in \citet{Li:2020} (see their Fig. 2).  Here we give a short description.

At each radial location, we calculate the composition and proportions of gaseous and condensed (dust) phases using the GRAINS code.  A small portion of the dust, called the ``decoupled dust'', is isolated from the system to form planetesimals.  The timescale for dust/planetesimals to decouple from the disk (so that it no longer advects with the gas or interacts with it dynamically) is $t_{\rm dec}=1.5\times10^4$ years.  The value of  $t_{\rm dec}$ is crucial for the final solid composition \citep{Li:2020}. Since we don't have a reliable physical model to calculate $t_{\rm dec}$, we choose this relatively long timescale to be consistent with \cite{Li:2020} where we found good agreement with observations.  The remaining dust, called ``advected dust'', remains coupled with gas and flows to neighbouring regions.  At that point a new chemical equilibrium is calculated at each radial location for the coupled dust and gas. This process is repeated until the end of disk evolution.  

In this work, we do not consider effects such as aerodynamic drag during the period of time that planetesimals grow and decouple from the gas.  The effect of this drag would be to drive the condensed material somewhat closer to the host star. We also ignore the effect that large decoupled planetesimals (e.g. km size) may not fully evaporate during the outburst. Since we do not yet follow the size distribution of the decoupled solids (which form planetesimals), we assume all dust, including coupled and decoupled, evaporates completely during the previous outburst for simplicity.  Thus, we start the chemistry calculation after the last outburst.

As the chemical calculation is the most expensive part of these simulations, during the outburst the disk chemistry is updated every 10 years.  After the outburst, it is updated every 100 years.  Tests similar to those in \citep{Li:2020} show that this frequency is sufficient to keep an accuracy better than 1\% of the final results.

\section{Results}\label{sec:results}

All initial conditions for the MCCs we use in this paper have a mass of $M_{\rm C}=1 M_\odot$, temperature of 15 K, and angular velocity of $1\times 10^{-14} \rm\ s^{-1}$.  These values are consistent with observations of MCCs \citep{Jijina:1999,Goodman:1993}.  We compare our outburst model results with the results of a disk that has a constant $\alpha$ viscosity of $\alpha=1\times 10^{-3}$.

\subsection{Outbursts and disk evolution}

\begin{figure*}
\includegraphics[width=0.5\textwidth]{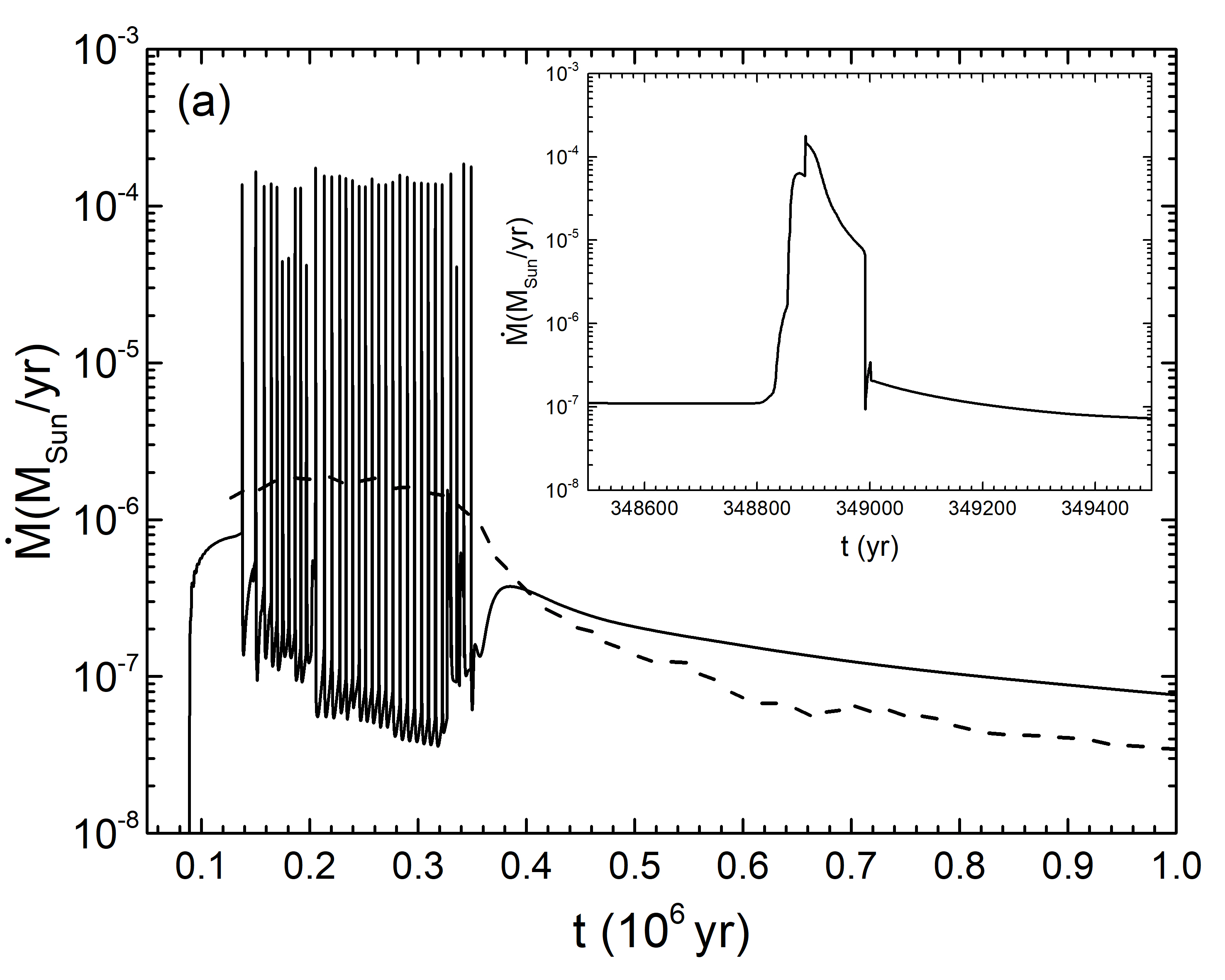}
\includegraphics[width=0.5\textwidth]{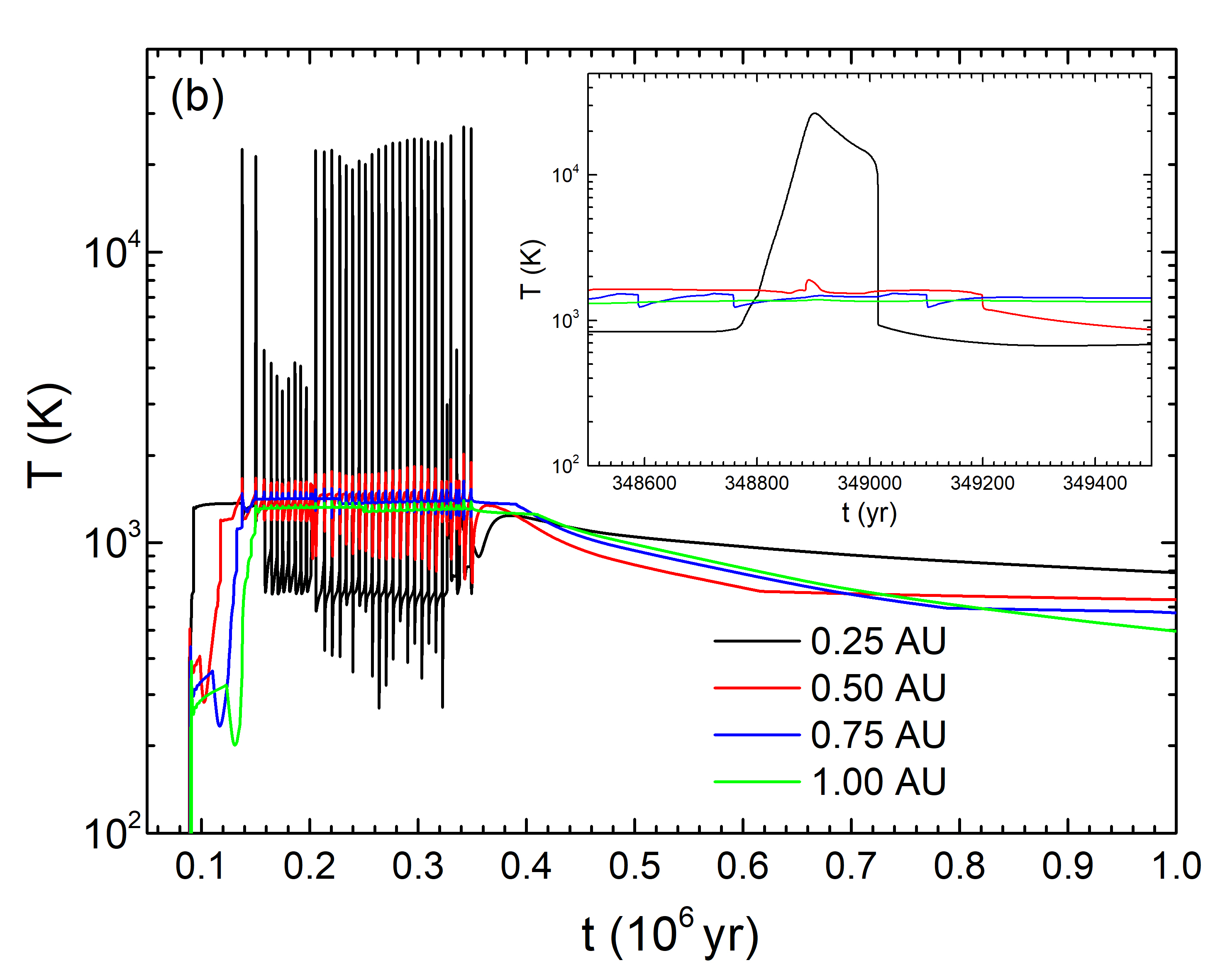}
\includegraphics[width=0.5\textwidth]{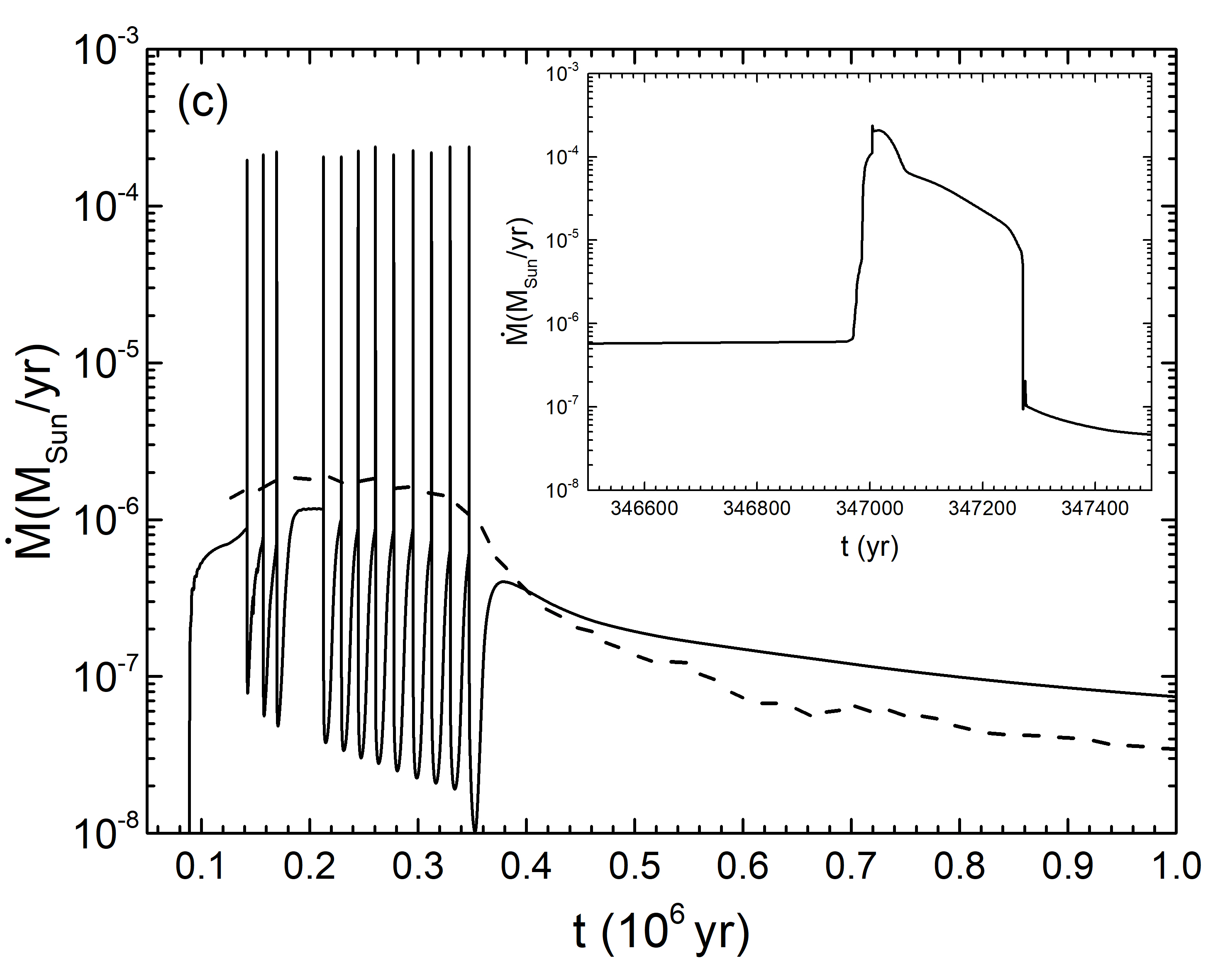}
\includegraphics[width=0.5\textwidth]{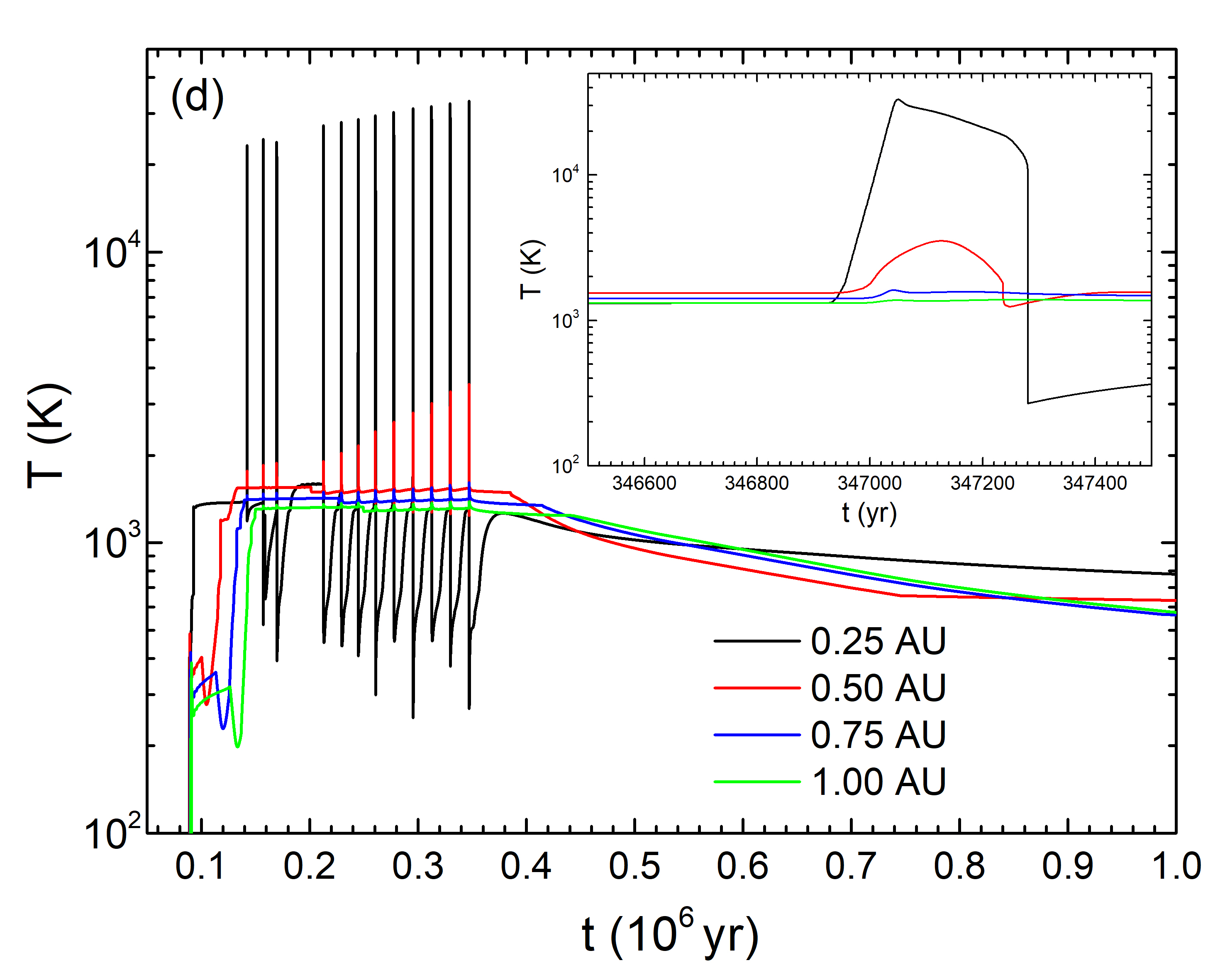}
\caption{
(a) Mass accretion rate and (b) midplane temperature for four different radial distances versus time for the default outbursting disk.  The dashed line in panel (a) shows the mass accretion rate for the constant-$\alpha$ disk model.  The inserts show the accretion rate and midplane temperature during the final outburst.  Here the initial mass, temperature, and angular velocity of the molecular cloud core are 1 $M_{\rm Sun}$, 15 K, and $1\times 10^{-14} \ \rm s^{-1}$, respectively.  (c) and (d) are mass accretion rate and midplane temperature for high ionization temperature model.
    }
    \label{fig:outbursts}
\end{figure*}

The default outburst model produces around 30 episodes of high accretion, with associated bursts of luminosity from the central star.  Figure \ref{fig:outbursts} shows the accretion rate as a function of time for the first million years of the disk evolution.  Each outburst lasts roughly 200 years as shown in the insert of Figure \ref{fig:outbursts}.  We focus specifically on the physical evolution of the disk beginning just prior to the final outburst, and the chemical evolution starting at the time when the midplane temperature at 1 AU peaks.  The high ionization temperature model shows fewer episodes and a longer period of outbursts.

\begin{figure}
	\includegraphics[width=1.0\columnwidth]{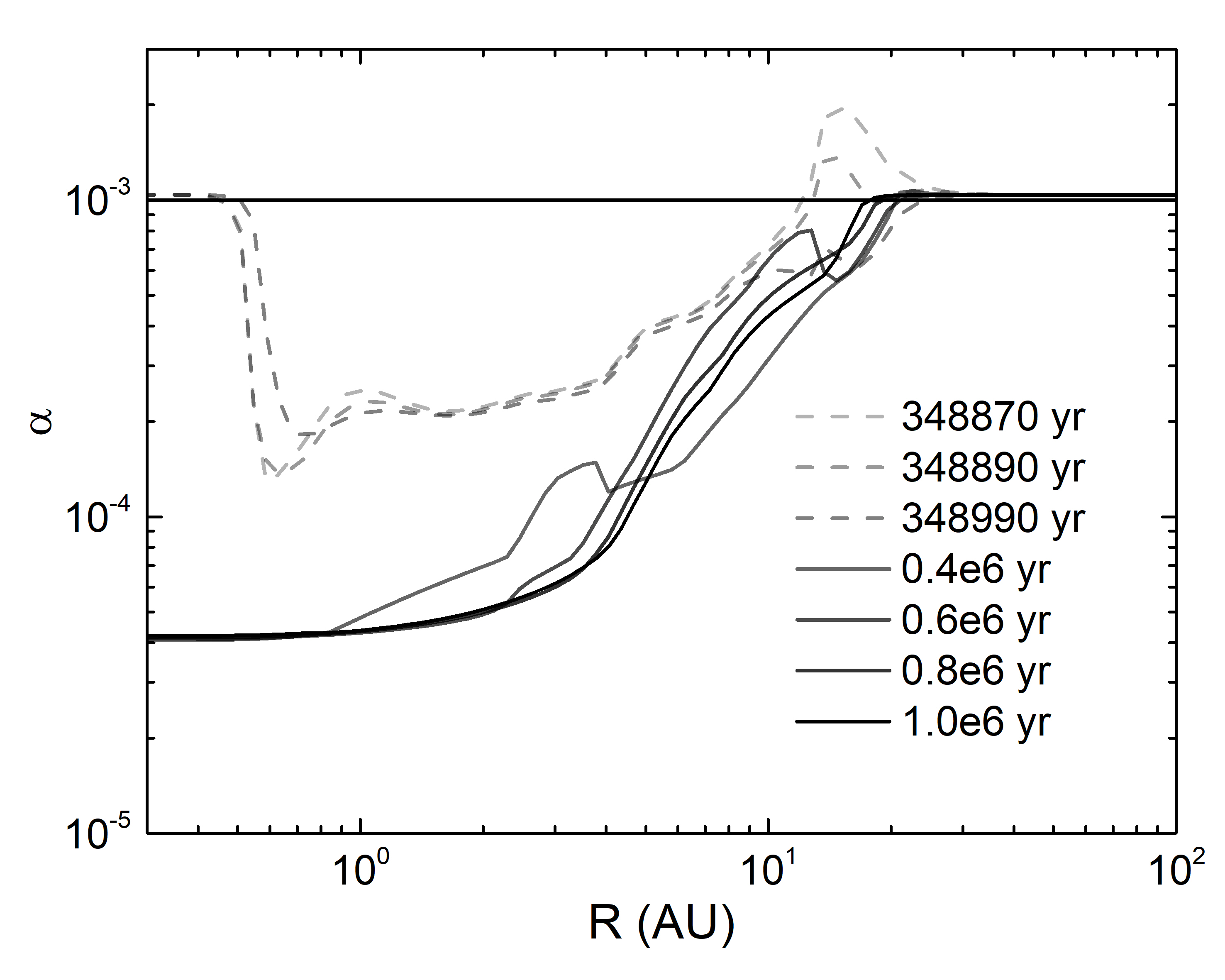}
	\includegraphics[width=1.0\columnwidth]{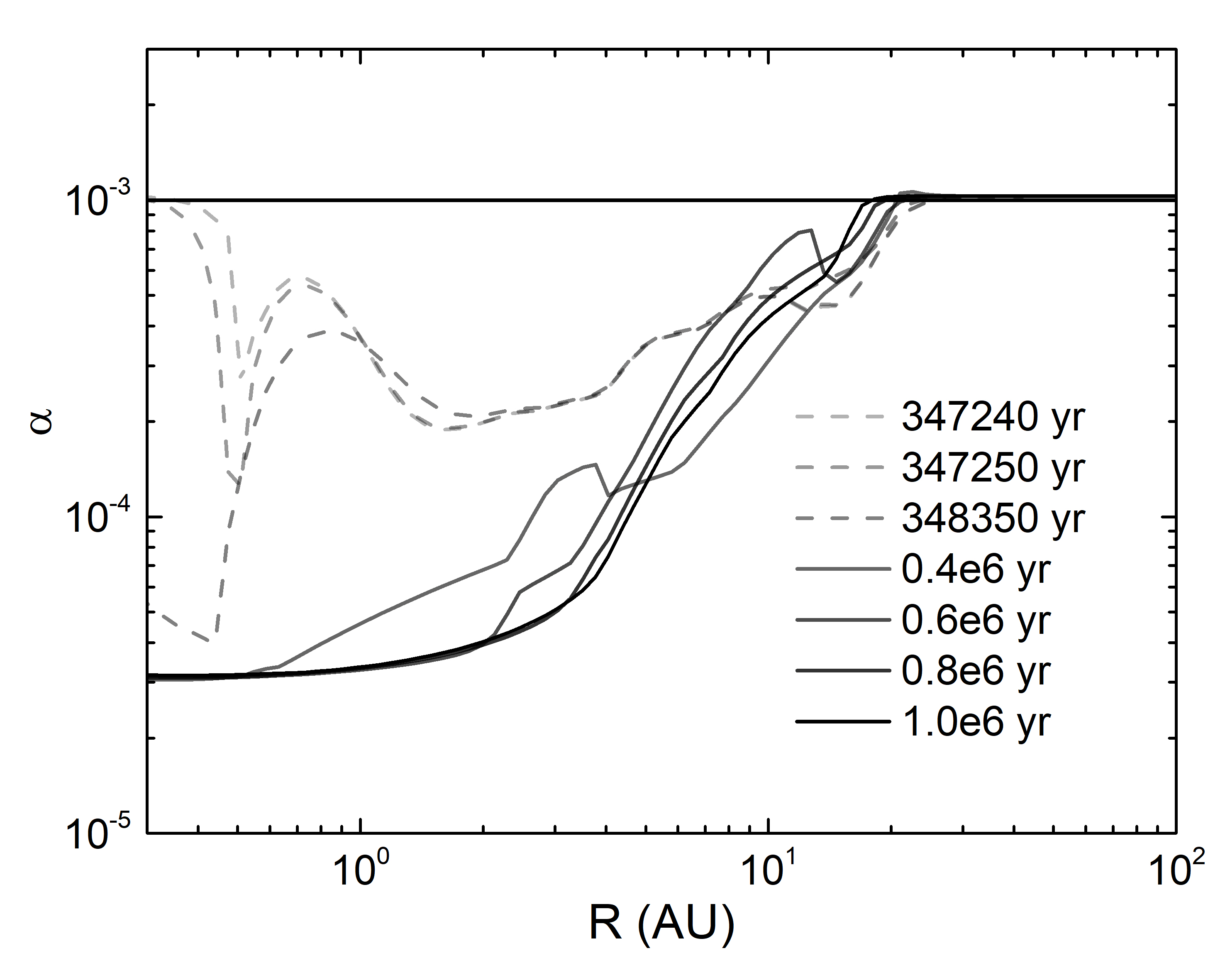}
    \caption{
    $\alpha$ values for the  default outburst model and constant alpha model (top panel).  The constant $\alpha$ is the solid line at $10^{-3}$.  The viscosity for the outbursting disk evolves with time as shown.  The dashed lines correspond to times during an outburst, while solid lines are the subsequent evolution.  The initial molecular cloud core parameters are the same model as in Figure \ref{fig:outbursts} (1 $M_{\rm Sun}$, 15 K, and $1\times 10^{-14} \ \rm s^{-1}$). We also show the results for the high ionization temperature model (bottom panel).
    }
    \label{fig:alpha}
\end{figure}

\begin{figure*}
\gridline{
    \fig{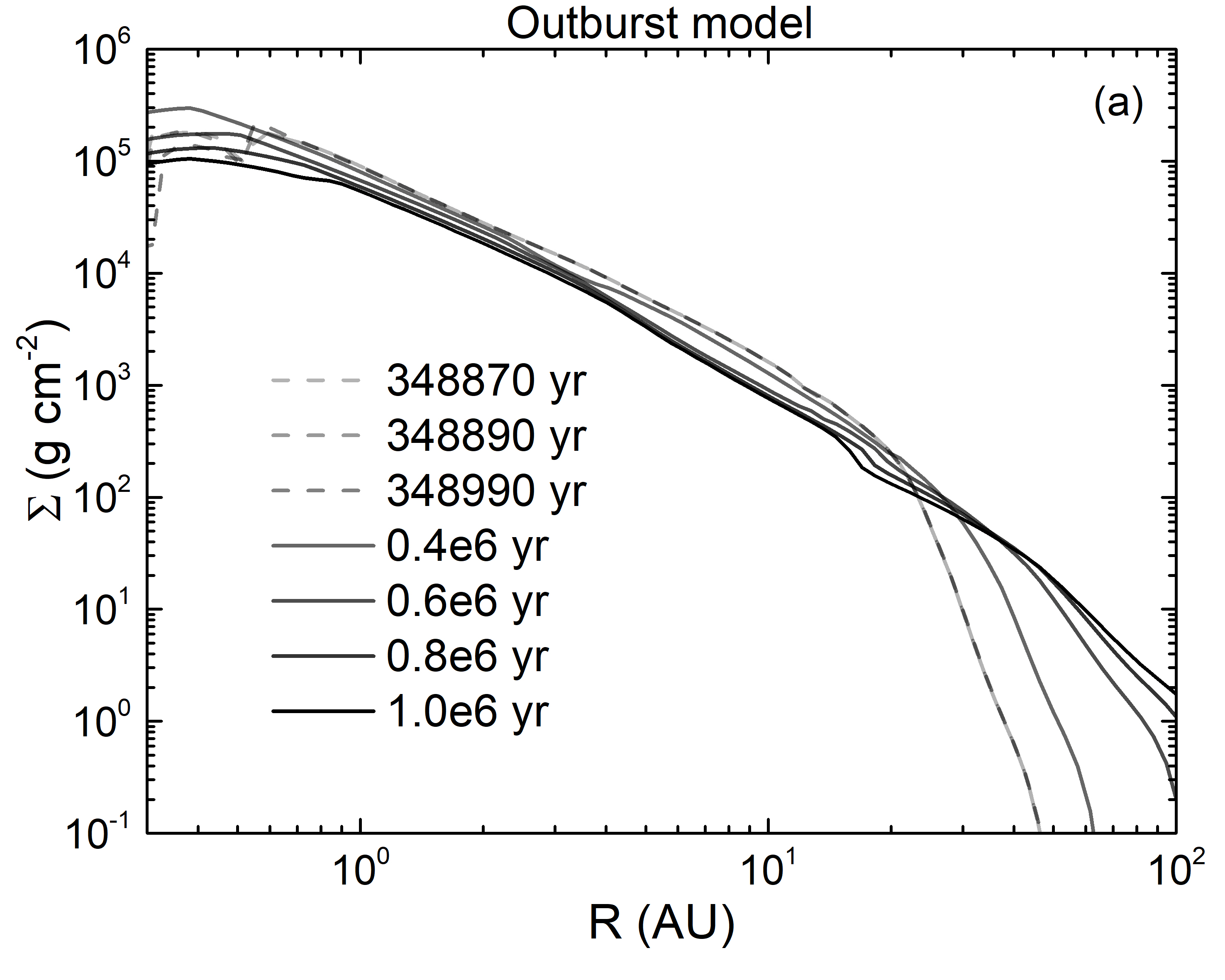}{0.4\textwidth}{}
    \fig{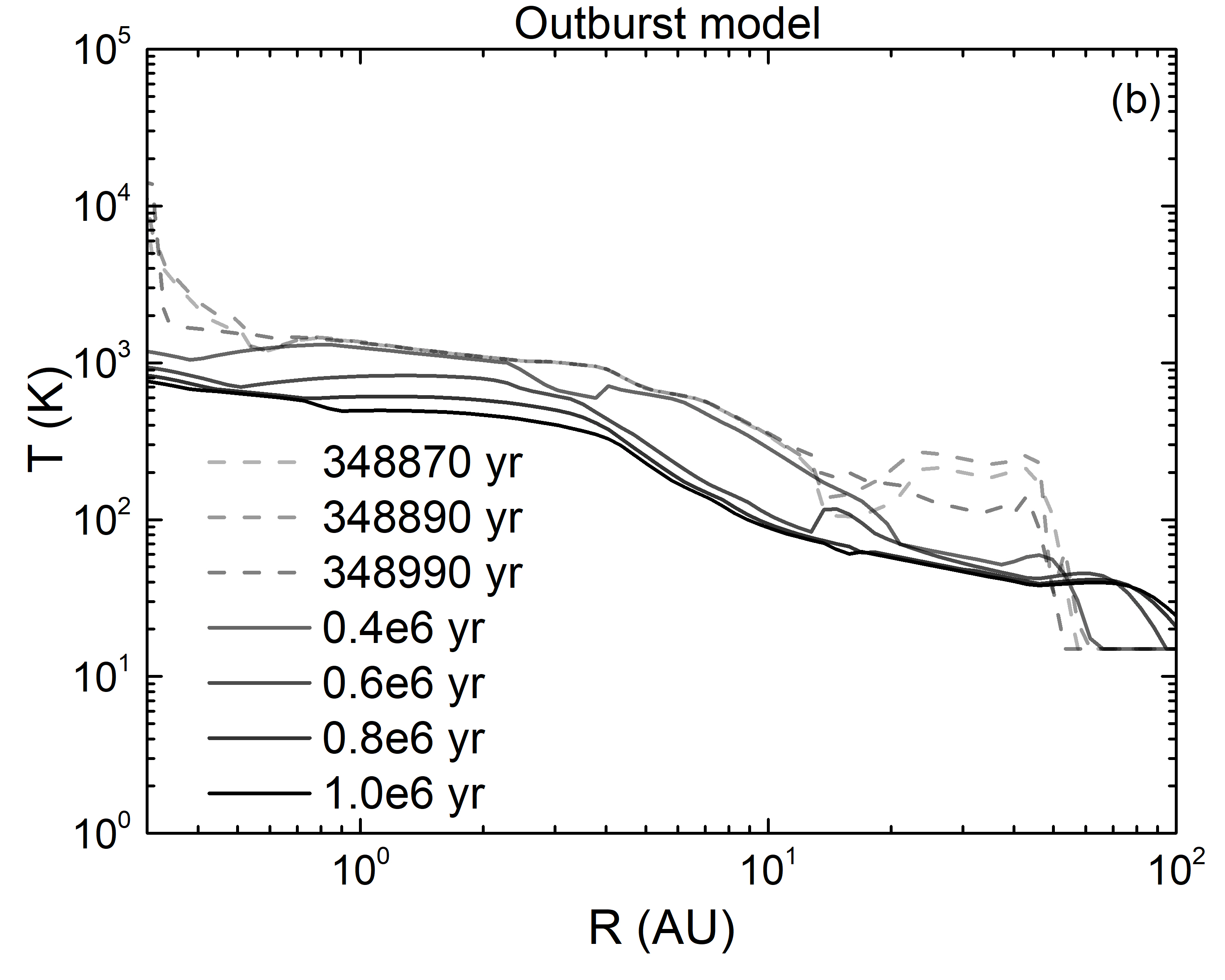}{0.4\textwidth}{}
}
\gridline{
    \fig{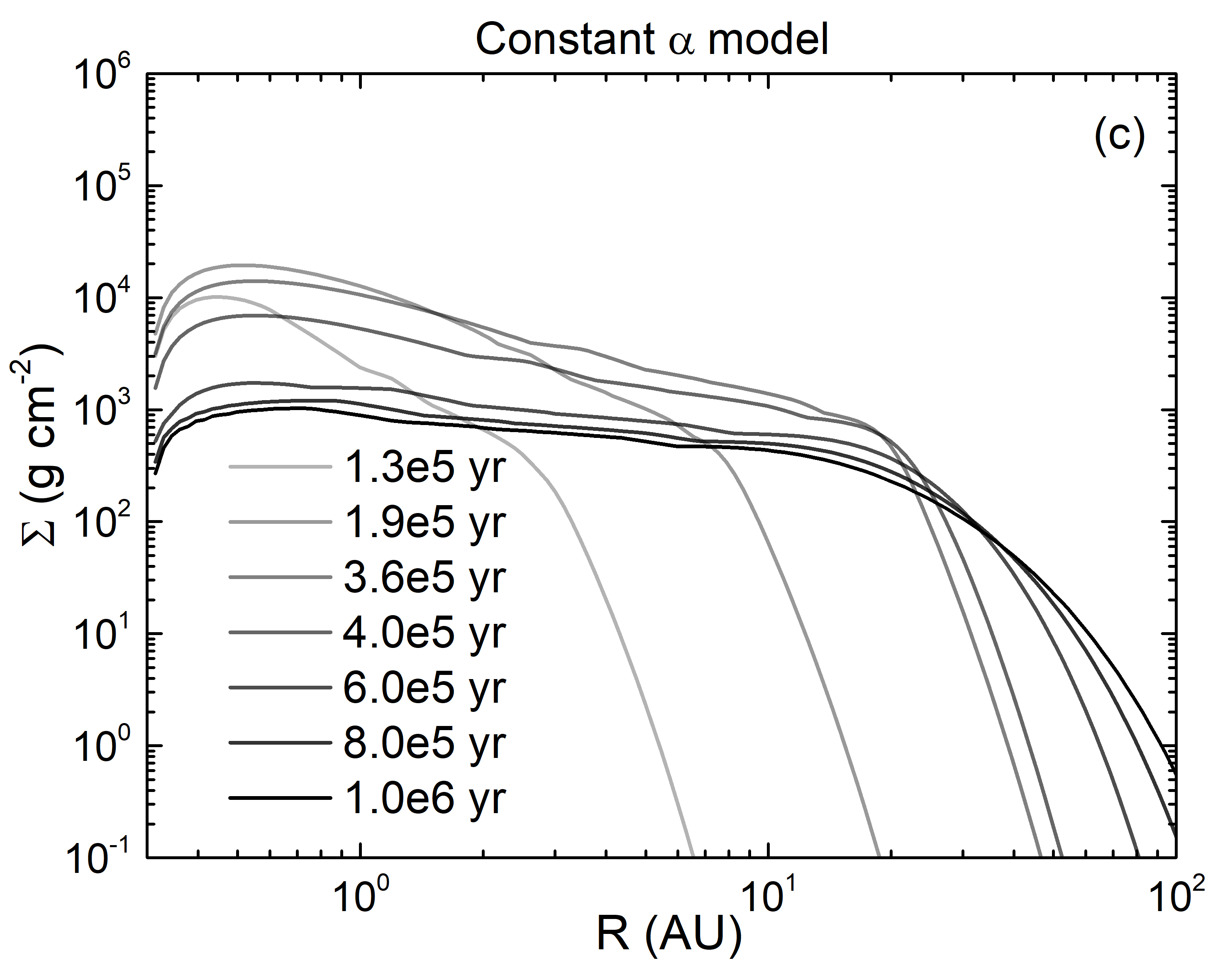}{0.4\textwidth}{}
    \fig{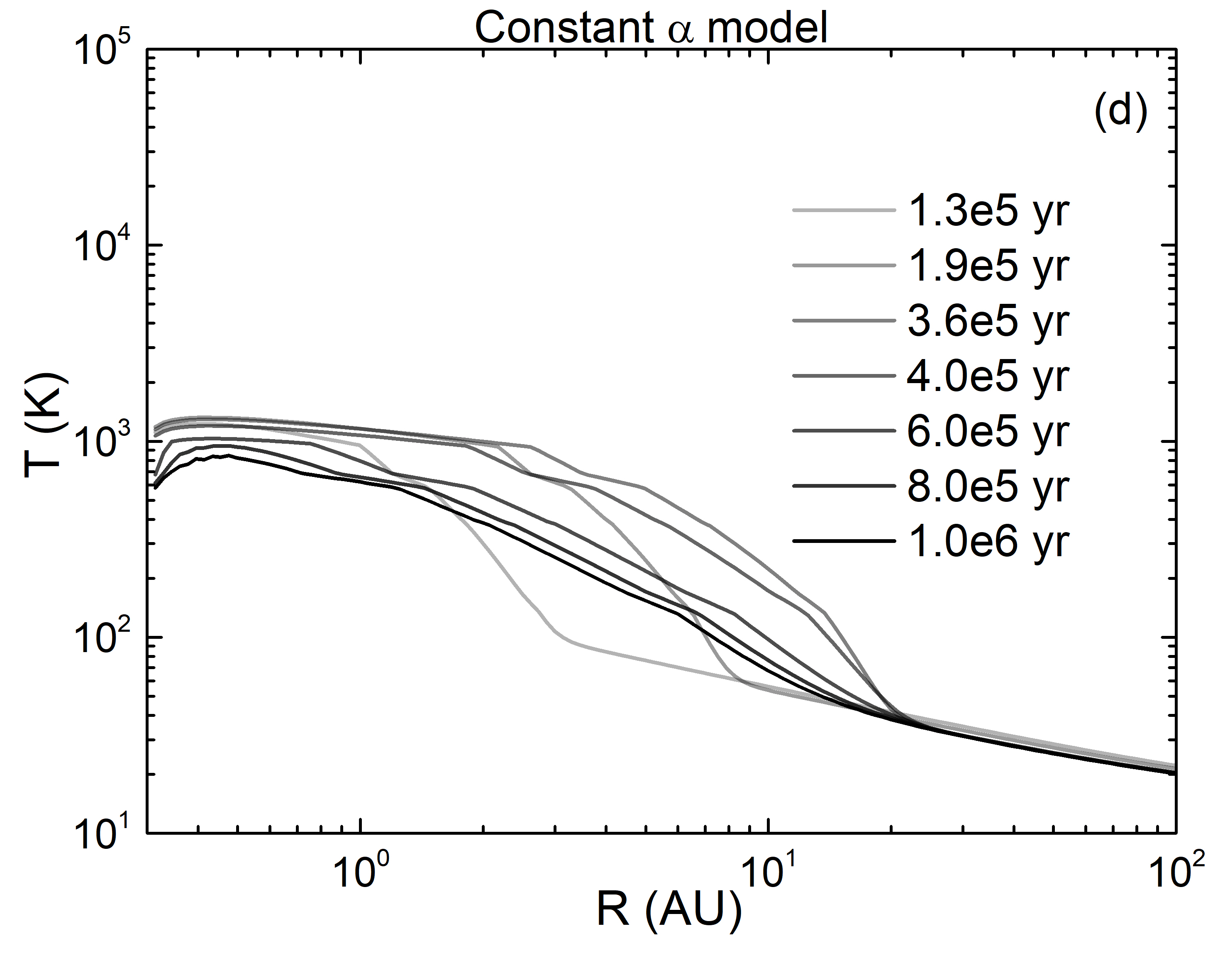}{0.4\textwidth}{}
}
\gridline{
    \fig{sur1800Ke.jpg}{0.4\textwidth}{}
    \fig{tem1800Kf.jpg}{0.4\textwidth}{}
}
\caption{
 Top: Default outburst model. Surface density and midplane temperature evolution of an outbursting disk.  The disk forms from a molecular cloud core with initial mass 1 $M_{\rm \odot}$, temperature 15 K, angular velocity  $1\times 10^{-14} \ \rm s^{-1}$, and $T_{\rm critical}=1500$ K.  The dashed lines correspond to the evolution during the final outburst.   Middle: The same information but using a constant $\alpha$ model for the disk where $\alpha=1\times 10^{-3}$.  Note that the temperatures for the constant $\alpha$ model are never sufficient to vaporize refractory elements $\sim 1500$ K, while the outbursting disk reaches temperatures in excess of 10,000 K in the inner AU, and up to nearly 1500 K at 1 AU.
 Bottom: The same information for the high ionization temperature model.}
\label{fig:diskproperties}
\end{figure*}

Unlike the constant-$\alpha$ model, the viscosity of our outbursting disk changes throughout its evolution.  Figure \ref{fig:alpha} shows how the viscosity varies as a function of time and distance from the central star.  These changes are driven by the rapid increase in viscosity during the outburst, which causes a sudden increase in temperature in the disk midplane of the affected region.  The effects of the temperature change are strongest in the inner AU, with additional heating occurring out to a few AU.  Figure \ref{fig:diskproperties} compares the evolution of the disk surface density and midplane temperature as a function of radial distance for the outburst models and for the constant-$\alpha$ model.

\begin{figure}
	\includegraphics[width=1.0\columnwidth]{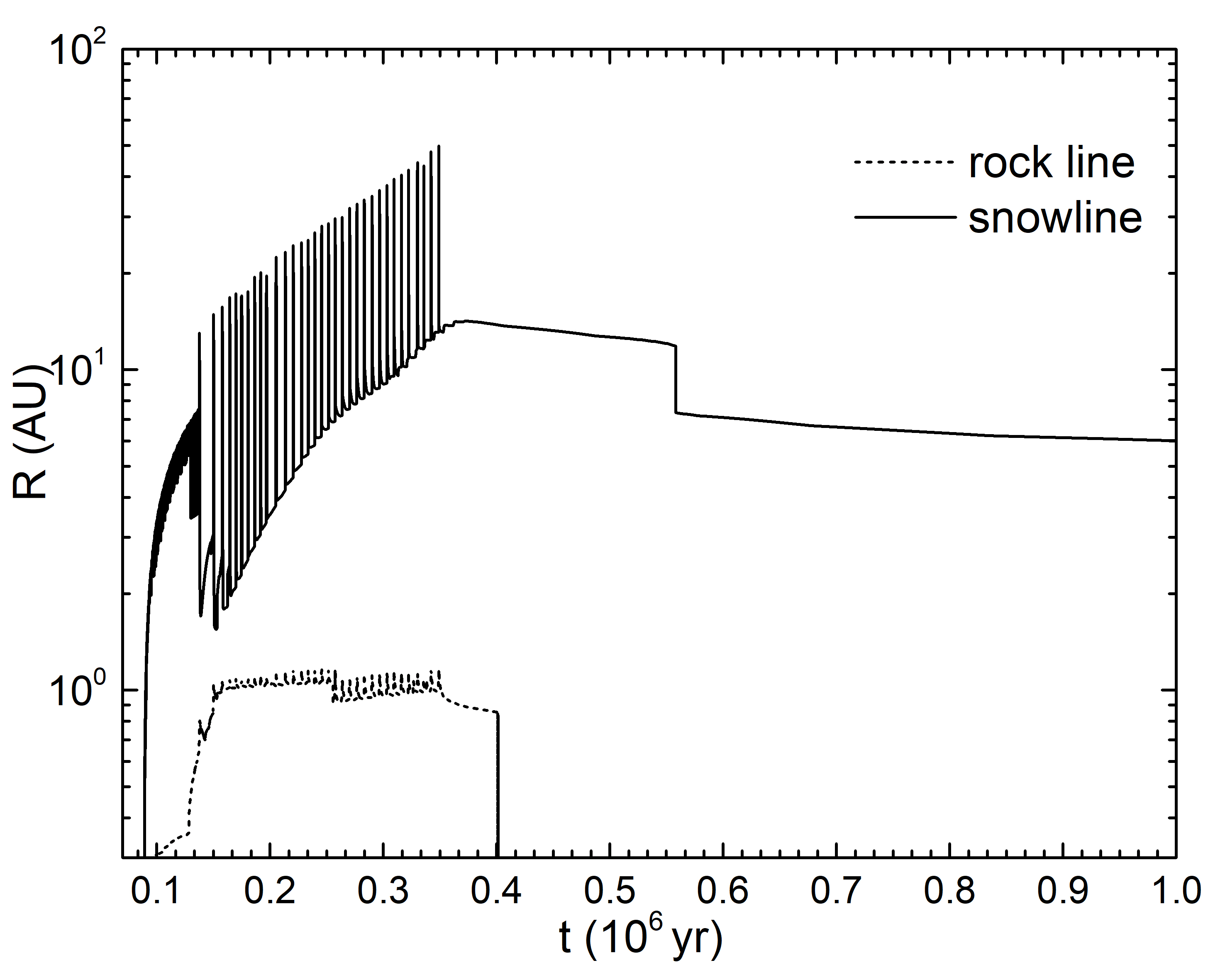}
	\includegraphics[width=1.0\columnwidth]{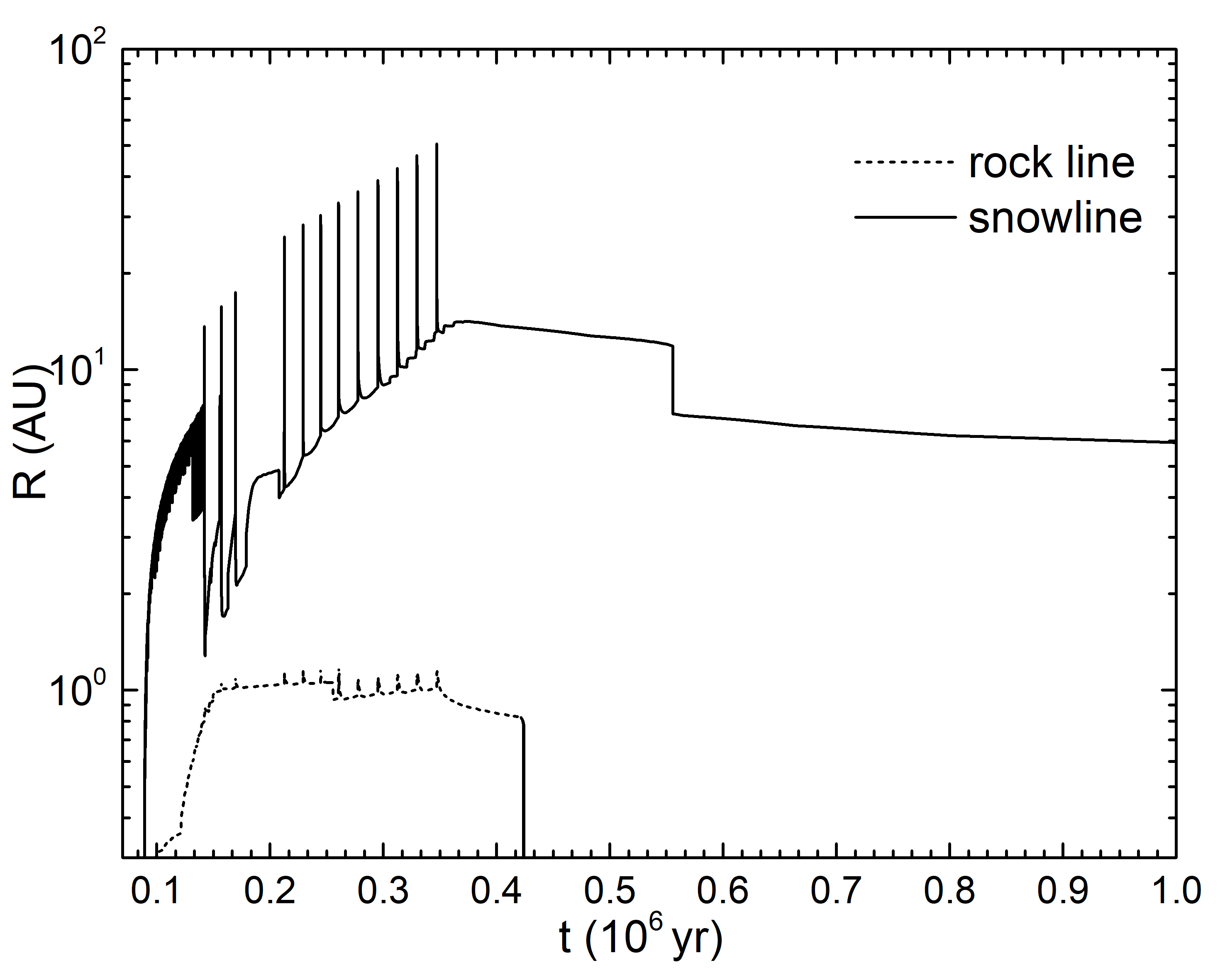}
    \caption{
    Top: The radial evolution of the rock line and snowline over time for the default outbursting disk. The snowline is the radius at which the midplane temperature is 170 K while the rock line is 1300 K---the 50\% condensation temperature of Si \citep{Li:2020}. Bottom: Same information for the high ionization temperature model.
    }
    \label{fig:rockline}
\end{figure}

Because viscosity is the primary mechanism for angular momentum transfer, the lower nominal viscosity of the outburst model causes an increase in the surface density of the disk, by roughly a factor of 10 following the outburst episodes for the inner few AU of the disk.  In the middle portion of the disk, near 10 AU, the surface density is lower for the outburst model.  At large distances near 100 AU, the surface density of the outburst model again rises above that of the constant-$\alpha$ model.  During an outburst, however, the surface density of the inner AU drops significantly as disk material is quickly accreted to the central star.

Following the outbursting phase, both models have similar temperature profiles (Figure \ref{fig:diskproperties}).  At 1 Myr, the temperatures peak in the inner disk at roughly 800 K, and decline as a function of distance as the entire disk cools.  During the outburst, the temperature of the inner AU (within 0.3 AU) of the disk rises by more than a factor of 10 to $T \gtrsim 10,000$ K over a relatively brief timescale of a few decades, indicating that the thermal instability occurred in the innermost regions of the disk \citep{Kadam2020}.  These temperatures are sufficient to vaporize any condensed material in the disk.  The temperature just beyond the inner AU (at around 0.5 AU) also increases, by roughly a factor of two---from 1000 K to 2000 K.  Even these modest temperatures are sufficient to vaporize most condensed phases.  On the other hand, in the constant-$\alpha$ model the midplane temperature remains below the condensation temperatures of many refractory elements.

The effects of the outbursts on dust formation are shown in Figure \ref{fig:rockline} where the evolving locations of the rock line and snow line are plotted over time.  The rock line, where half of Silicon is in the condensed phase, is less than $\sim$ 1 AU for the first 150,000 yr, and can be pushed to $\sim$ 1.15 AU during outbursts.  The snowline, by contrast, can be driven beyond 10 AU during outbursts.  This result is consistent with previous ALMA observations on FUors \citep{Cieza:2016} and theoretical calculations \citep{Martin:2012, Vorobyov:2022}.

\subsection{Chemical evolution and abundance patterns}

\begin{figure}
    \includegraphics[width=0.8\columnwidth]{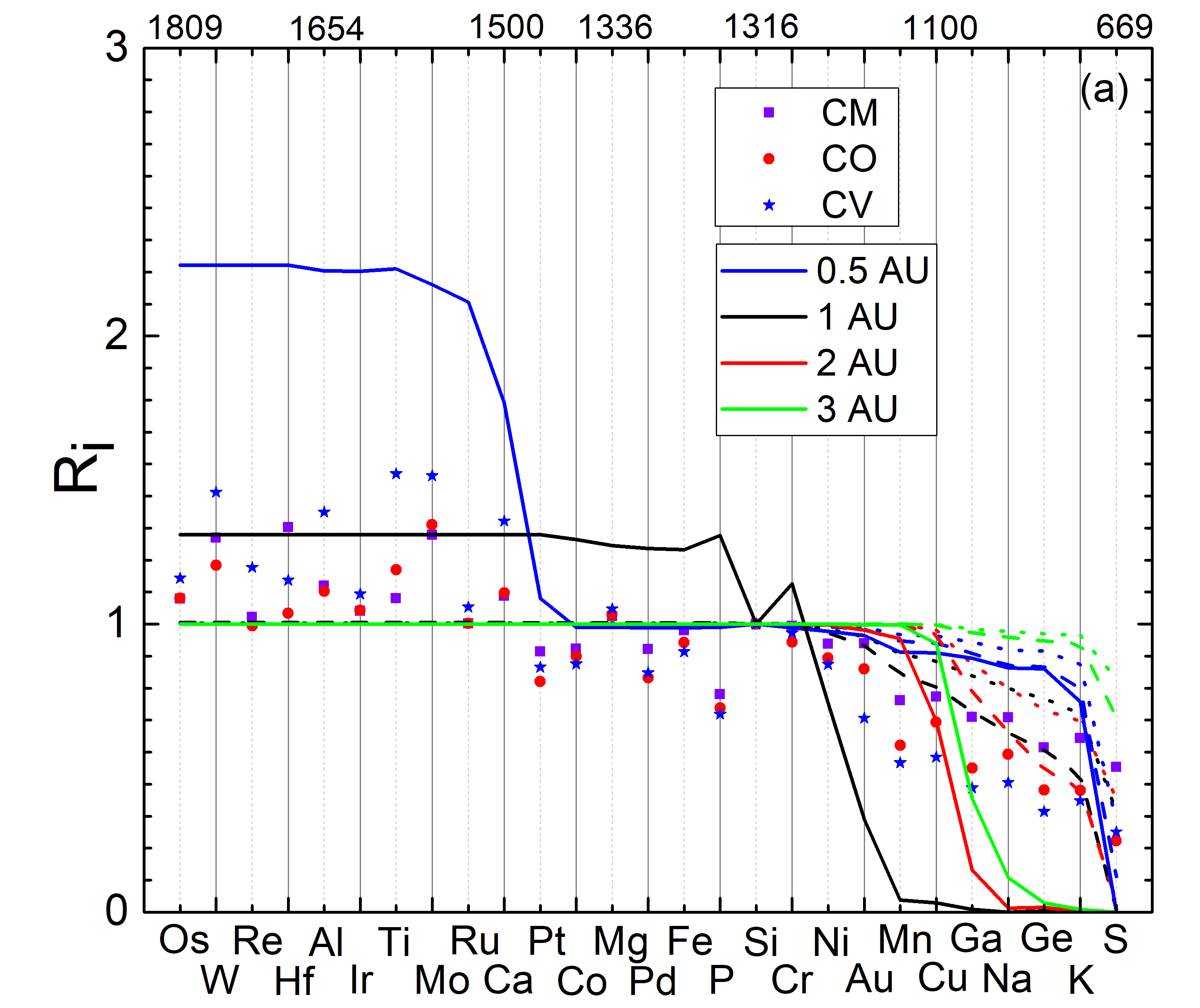}
    \includegraphics[width=0.8\columnwidth]{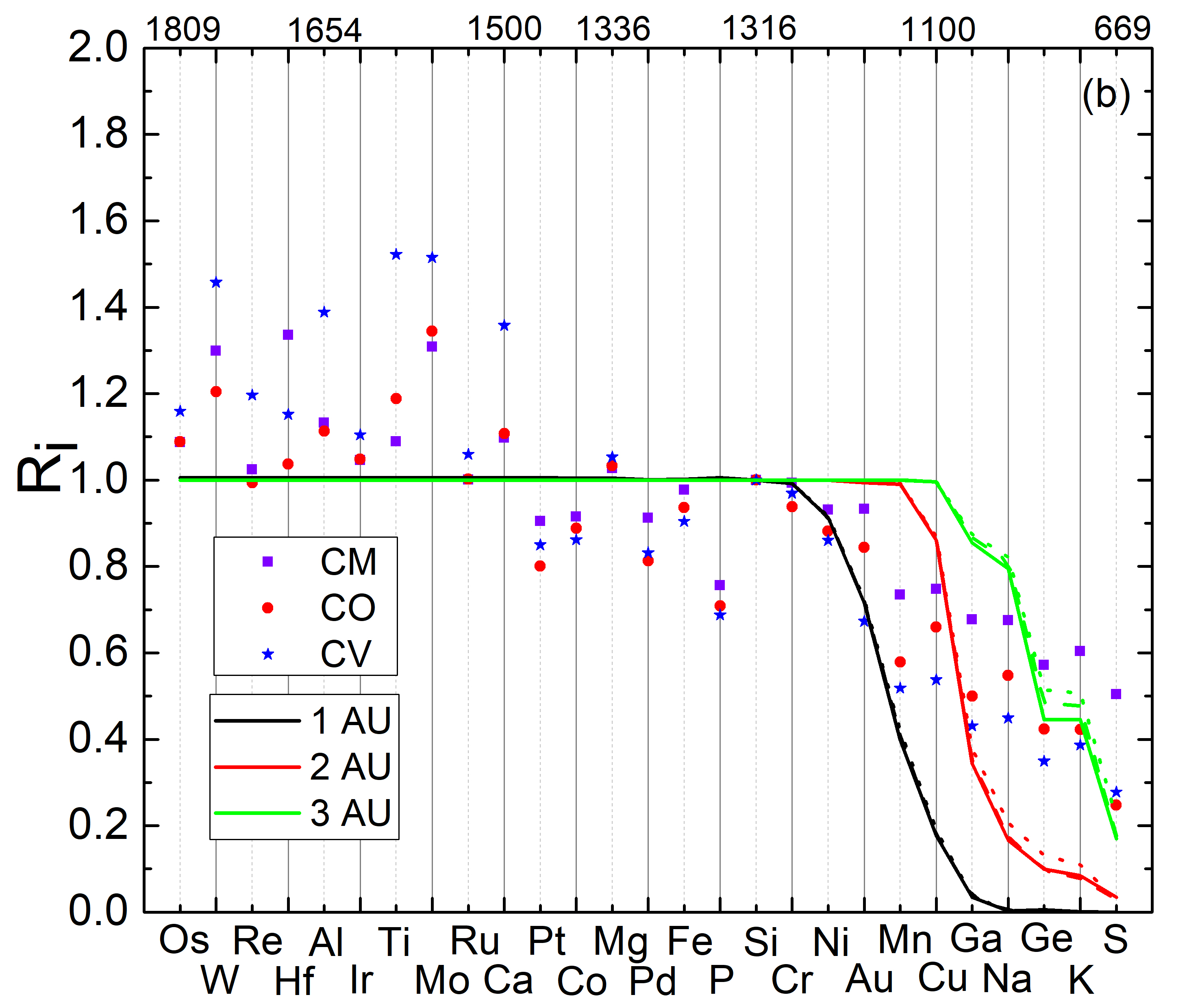}
    \includegraphics[width=0.8\columnwidth]{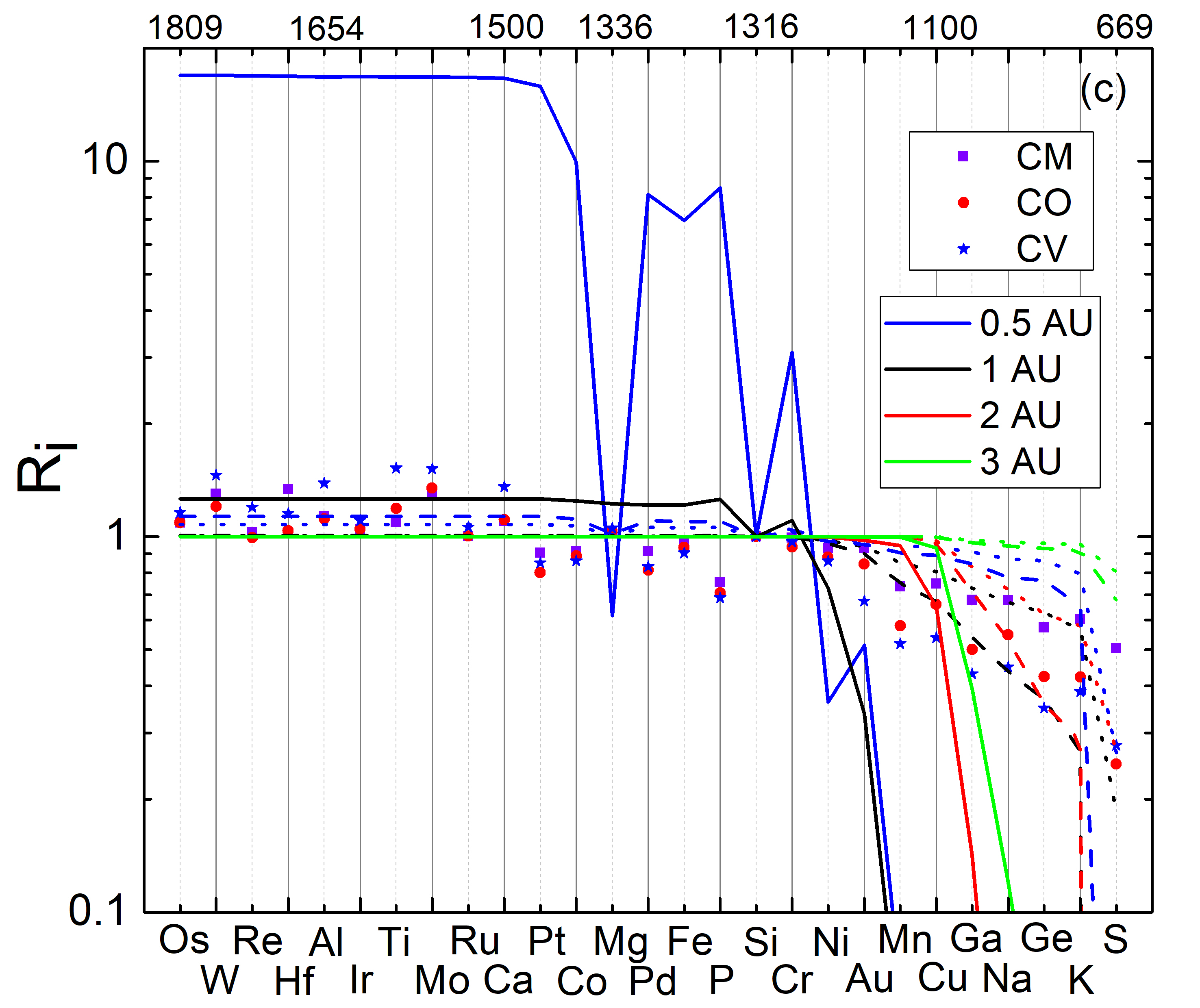}
    \caption{
     Final elemental abundances in planet-forming materials from (a) our default outburst model, (b) the constant $\alpha$ disk model, and (c) high ionization temperature model for four different distances and for three separate times.  Measured abundances ($R_i$ = (x/Si)$_{\rm sample}$/(x/Si)$_{\rm CI}$) for these elements for different chondrites are also shown.  The 50\% condensation temperatures of a few elements are displayed across the top axis.  The outburst model yields an excess of refractory elements and matches the volatile depletion profile---especially at a distance of 1 AU.  The constant $\alpha$ model does not produce any excess of refractory materials and is a poor match, relative to the outburst model, for the depletion pattern of volatile elements. The times for the solid, dashed, and dotted lines are 0.37 Myr, 0.7 Myr, and 1 Myr, respectively.
     }
    \label{fig:ri}
\end{figure}

Because of the low disk temperatures in the constant-$\alpha$ model, all refractory and the majority of moderately volatile elements are locked in the condensed phases so that the disk evolution does not fractionate them from each other---the gaseous phase contains only volatile elements during the entire evolution of the inner disk.  The different flow rates between the gaseous and condensed phases in the disk change the relative proportions of volatile to refractory and moderately volatile elements, leading to the abrupt drop of the moderately volatile element depletion factors in the modeled condensates (Figure \ref{fig:ri}b).  This abrupt drop does not match the steadily declining element patterns observed in carbonaceous chondrites (Figure \ref{fig:ri}).

The situation in the inner disk is significantly different for the outburst model.  The outbursts heat the inner disk to over ten thousand Kelvins (Figure \ref{fig:diskproperties}), which is hot enough to evaporate all condensed materials in the inner half-AU of the disk.  As the disk cools, partial condensation of refractory elements, coupled with the differential flow of gaseous and condensed phases, is able to fractionate the moderately volatile from refractory elements, and to reproduce the observed element patterns.  This result is similar to results from the disk model presented by \cite{Cassen:1996} and \cite{Li:2020}, which starts with a hot disk at a prescribed initial temperature.

Figure \ref{fig:ri} compares the resulting depletion factors of 27 chemical elements, predicted by the outburst model and by the constant-$\alpha$ model, at four different distances and three different times in the evolution of the disk.  Also shown for reference are the approximate 50\% condensation temperatures for several elements (taken from \citet{Li:2020}).  Of particular importance is that the constant-$\alpha$ model does not produce the excess refractory abundances (normalized to silicon), and the volatile depletion is significantly steeper than observed in chondrites.  Meanwhile, the outburst model yields a refractory excess that is comparable to what is observed, and produces a depletion profile that is also a better match---especially for material that condenses near one AU.  Note that the formation of these chondrites is limited in both time and space. If disk accretion can last for a long time following an outburst, material at the inner disk can fall onto the star so that the excess of refractory elements diminishes with time.  Snapshots of the individual chemical abundances as a function of radius are shown for four different times in the appendix.

Because of the low disk temperatures in the constant-$\alpha$ model, which do not evaporate the refractory material, the relative abundances of the refractory elements match the initial composition of the MCC.  At the same time, the slow cooling of the disk keeps the most volatile elements in the gaseous phase while they advect with the H/He gas into the central star.  This advection causes an abrupt truncation of the volatile element depletion factors (as a function of condensation temperature) in the disk midplane since they have limited opportunity to bind into solid compounds.

The situation in the inner disk is different for the outburst model because nearly all of the condensed material evaporates in the inner AU of the disk, including the most refractory elements.  As the disk starts cooling again, the most refractory elements begin to condense, and the less refractory elements advect with the H/He gas toward the central star.  The excess of refractory materials in the dust arises because Si, the reference element, is only partially condensed and experiences some advection.  This results in higher relative abundances of more refractory elements because of normalization to Si.  The longer an element advects with the gas (determined by its condensation temperature) the less of it will contribute to the planet-forming material that condenses.

\section{Discussion}\label{sec:discussion}

In addition to matching the chondrite compositions, the early thermal evolution of the evolving disk (Figure \ref{fig:diskproperties}) in the outburst model also provides an explanation for some aspects of the formation of CAIs.  As the first solids that formed in the Solar System, mineralogical, elemental, and isotopic evidence requires CAIs and their precursors to have experienced multiple stages of evaporation and condensation \citep{MacPherson:2003,Petaev:2009b,Huang:2012}.  The outburst model provides the multiple heating events with subsequent condensation needed to explain these observations.  These heating events also yield the observed differences in formation times for CAIs vs chondrites as the CAIs would form during the outbursting phase (less than one million years) while the chondrites form as the disk cools following the final outburst.  Indeed, recent work by \citet{Forbes:2021}, suggests that global heating events, such as these outbursts, could play an important role in CAI formation.

A competing model to explain the CAIs, the x-wind model, suggests that CAIs formed within the x-region of the system (within a few solar radii of the Sun).  There, CAIs and their precursors are heated up by impulsive flares in the reconnection ring \citep{Shu:1977}.  This model faces some challenges, such as how solids condense at such short distances and how they are subsequently transported out to AU distances where chondrites form \citep{Desch:2010}.  In our outburst model, the gas accretion rates fluctuate by several orders of magnitude in the first 0.4 Myrs, causing large disk temperature fluctuations, from 1,000 K to up to a few times 10,000 K, within one AU (Figure \ref{fig:outbursts}).  In these regions, materials can be evaporated and re-condensed multiple times---allowing CAIs to form at distances closer to 1 AU (rather than 0.1 AU).

A second model to explain CAIs is nebular shocks \citep{Desch:2002}.  Sources of these shocks could include bow shocks from orbiting objects or gravitational instabilities in the protoplanetary disk \citep{Desch:2010}.  This model yields CAI-forming conditions at AU distances as does our outburst model.  One difference between these two models is the cooling timescales for the heated region.  Nebular shocks predict CAI cooling timescales of hours \citep{Desch:2002}, while the ambient material from stellar outbursts cool over decades to centuries.  Exploring the consequences of these differences in future work would give useful comparisons to observations.

Another issue where the outburst model may resolve existing tensions relates to carbonaceous (CC) and non-carbonaceous (NC) chondrites.  CC and NC groups of meteorites have different isotopic anomalies, with the CC group having more isotopes from neutron-rich processes \citep{Kruijer:2020}.  One explanation is that this difference indicates that CC and NC groups sampled different, isolated reservoirs within the solar nebula.  Specifically, during the collapse of the molecular cloud, early infalling material may have contained more neutron-rich isotopes than late infalling material.  If so, the nebular reservoir contributing to CC-group meteorites contains more early infall material than that of NC-group meteorites \citep{Kruijer:2020,Lichtenberg:2021,Johansen:2021}.  This model implies an isotopically heterogeneous parental molecular cloud.

The outburst model provides an alternative interpretation to the observed CC vs. NC isotopic differences, building on the model of selective destruction of isotopically anomalous presolar materials (related to their melting and evaporation points) \citep{Trinquier:2009,Koop:2018}.  Figure \ref{fig:diskproperties} shows that within the first Myr, the region between 1 and 10 AU is hot enough to partially evaporate some, but perhaps not all, isotopically anomalous presolar materials.  This would lead to isotopically different gaseous and condensed phases, which flow with different velocities as the disk evolves.  As a consequence, the inner disk would develop different isotopic anomalies compared to the outer disk, as shown in Figure 1 of \citet{Johansen:2021}, and chondrites that form at different locations in the disk would have different isotopic anomalies without requiring an initially heterogeneous molecular cloud.

\section{Conclusions}\label{sec:conclusion}

Stellar outbursts caused by the interplay between gravitational instability and magneto-rotational instability during star formation may resolve the stellar Luminosity Problem whereby stars are observed to accrete material too slowly to form within the disk lifetime.  Periodic bursts of accretion can provide the necessary material in sufficient time.  Here we showed that these bursts of accretion also produce radiative outbursts with sufficient energy to vaporize condensed material in the inner Solar System.  The subsequent temperature evolution of the disk, when coupled with calculations of both the chemical equilibrium and the dynamics of gaseous material, predicts the compositions of planet-forming materials that match CV, CO, and CM chondrites.  This model also has implications, and may provide a partial explanation, for other anomalies observed in Solar System material including CAI formation and differences between the CC and NC chondrites.

It appears that a number of properties of the Solar System materials can be explained by high temperatures within the protoplanetary disk.  However, a constant-$\alpha$ disk model is incapable of generating these temperatures.  Our outburst model produces the required temperatures.  This motivates further work to investigate the detailed implications of both the physics and the chemistry of this model, as well as its application to other Solar System observations for understanding the origin of planet-forming materials.

\medskip
\textbf{Acknowledgments: }
JHS, SH, and ZZ thank the National Science Foundation for their support for our work under grant AST-1910955.  MIP was supported by the NASA (80NSSC20K0346) and DOE-NNSA (DE-NA0003904) grants (PI - Stein B. Jacobsen). This work has been supported in part by the National Natural Science Foundation of China (NSFC) grant 12203018 and by a grant from Jilin Normal University.

\bibliography{outburstbib}{}

\appendix


\section{Snapshots for the abundance and disk properties}

In our simulation, we calculate the evolution of the disk from the collapse of molecular cloud cores. For the condensation of the elements, we begin our calculation near the time of the last outburst.  Here we show the $R_i$ values of all the elements, surface densities, temperatures, and pressures in the discs. $R_i$s for H, He, C, and N are set to be zero as their condensation temperatures are lower than the lowest temperature of equilibrium of our calculation.  We chose four snapshots: during the last outburst, just after the last outburst, around the mid time of the last outburst , and the end of the evolution.

\begin{figure}
    \includegraphics[width=1.0\columnwidth]{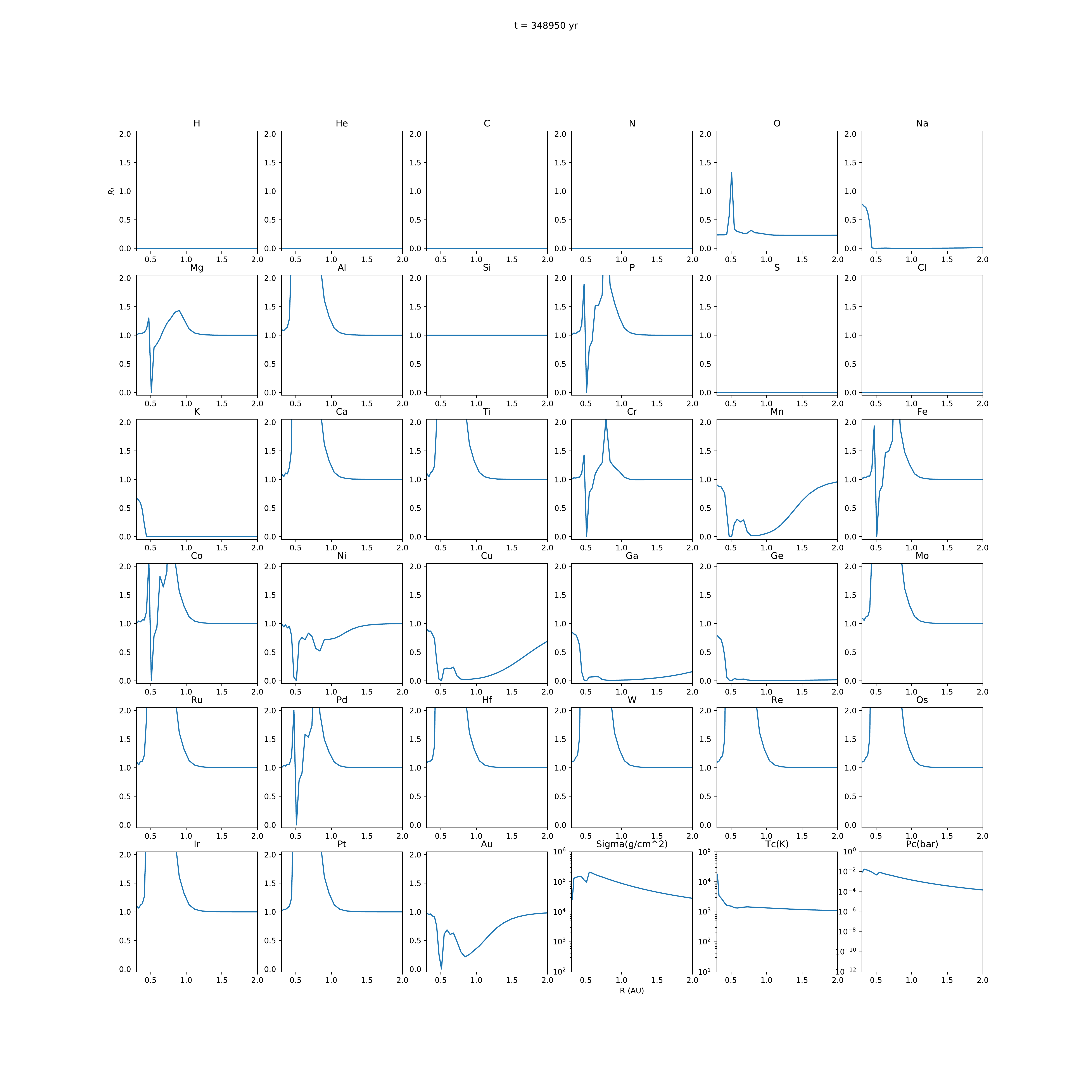}
    \caption{Snapshots for the abundance and disk properties at 348950 yr.
    }
    \label{fig:ri1}
\end{figure}

\begin{figure}
    \includegraphics[width=1.0\columnwidth]{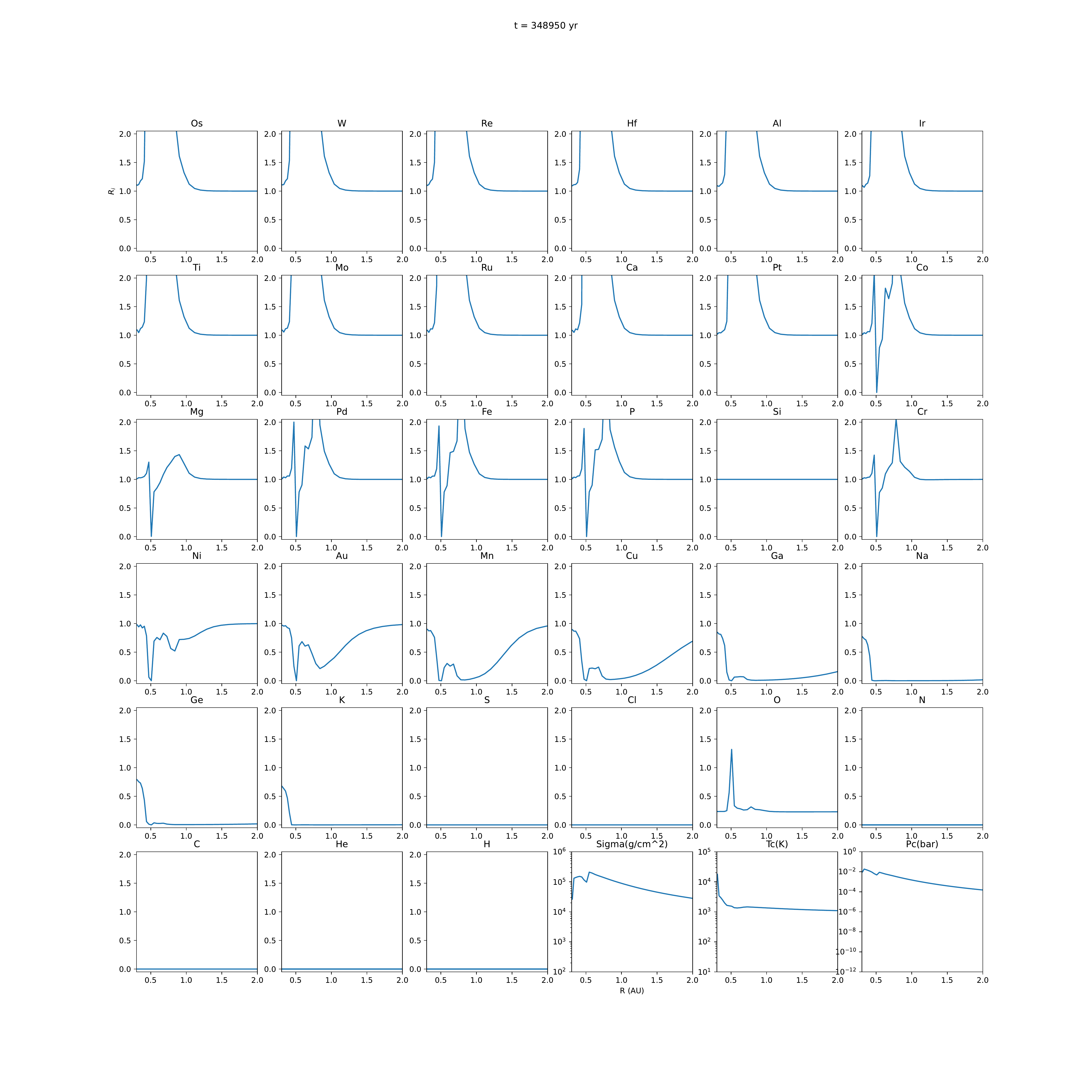}
    \caption{Snapshots for the abundance and disk properties at 348950 yr. The elements are ordered according to their T50s
    }
    \label{fig:ri1new}
\end{figure}

\begin{figure}
    \includegraphics[width=1.0\columnwidth]{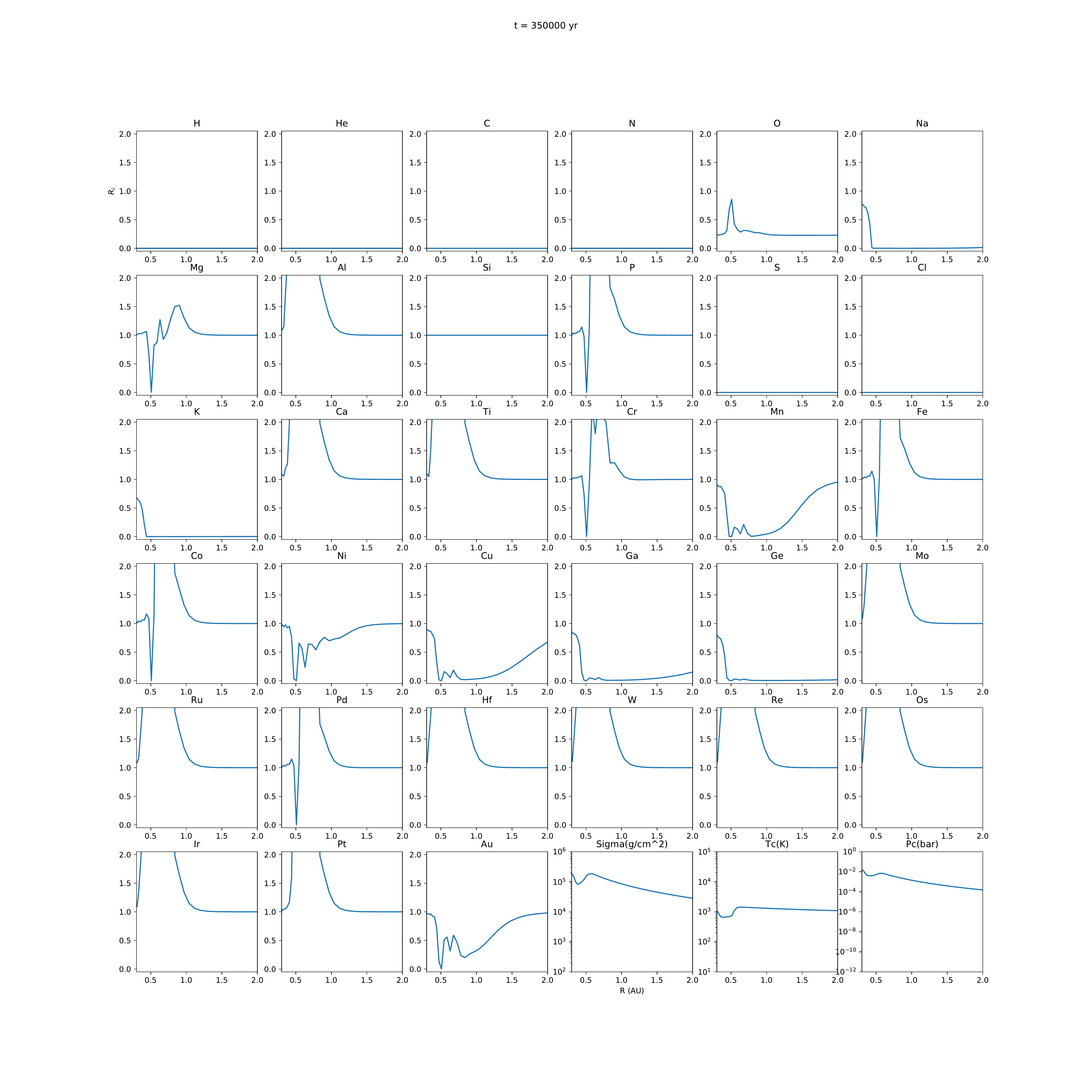}
    \caption{
    Snapshots for the abundance and disk properties at 0.35 Myr. 
    }
    \label{fig:ri2}
\end{figure}

\begin{figure}
    \includegraphics[width=1.0\columnwidth]{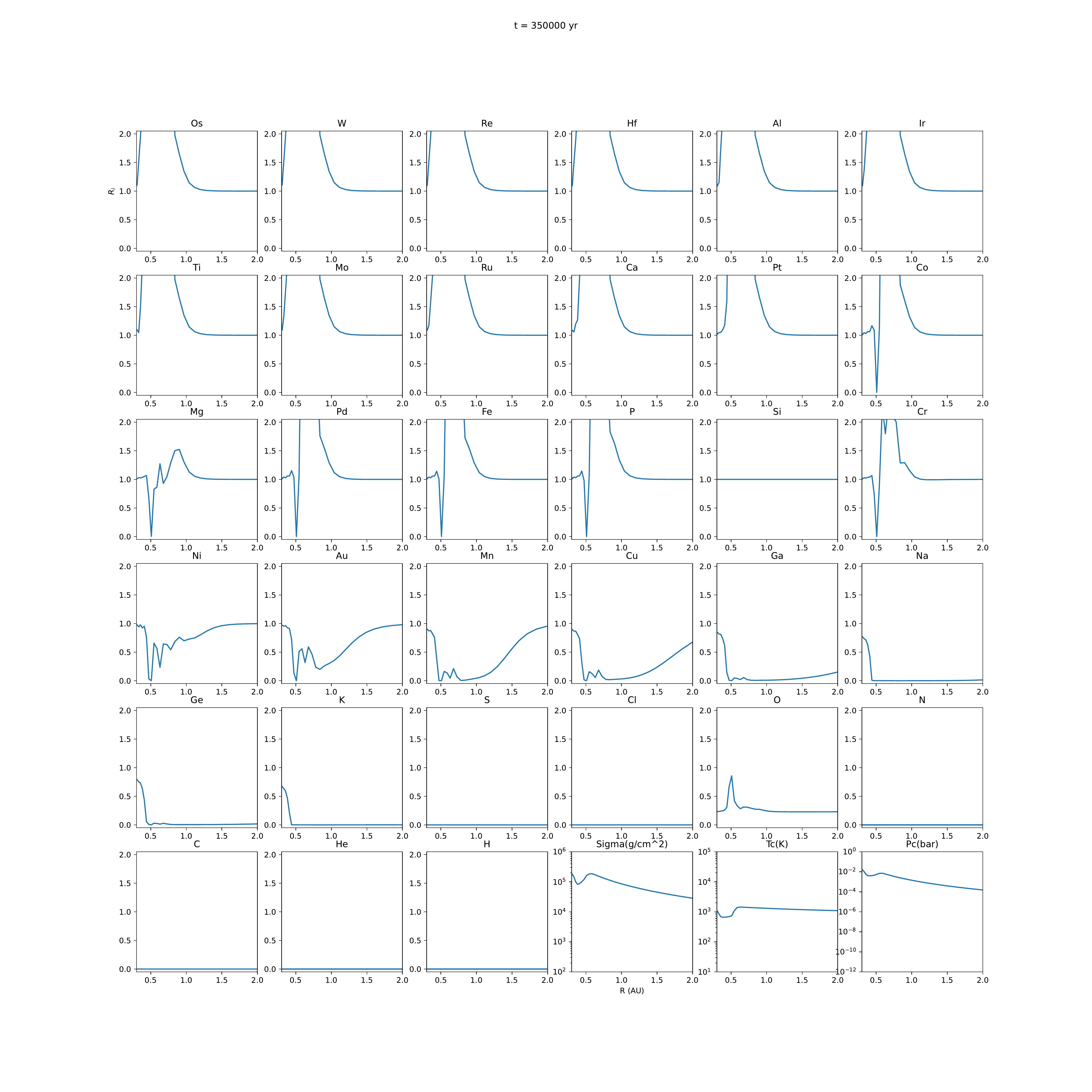}
    \caption{Snapshots for the abundance and disk properties at 348950 yr. The elements are ordered according to their T50s.
    }
    \label{fig:ri2new}
\end{figure}

\begin{figure}
    \includegraphics[width=1.0\columnwidth]{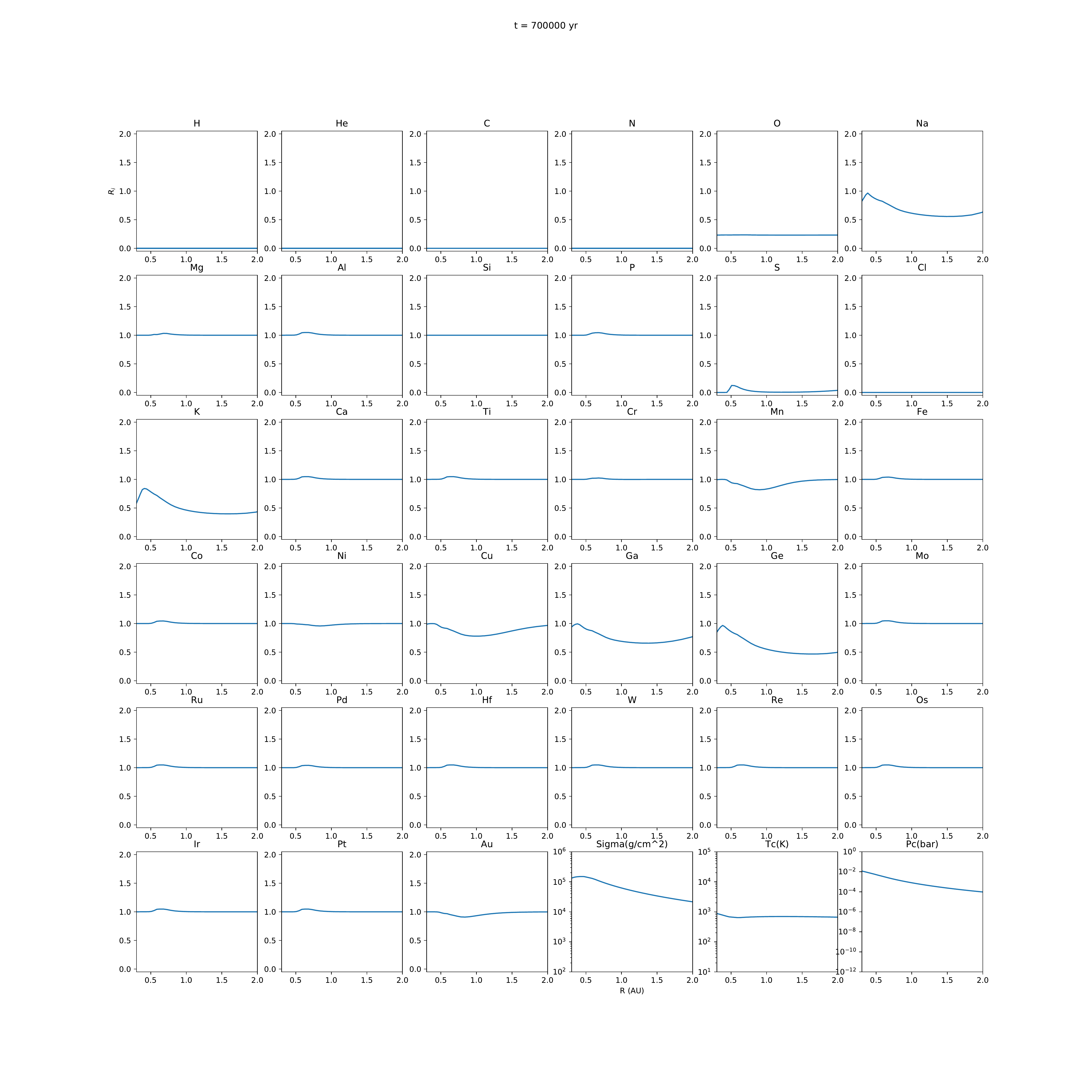}
    \caption{
    Snapshots for the abundance and disk properties at 0.7 Myr.
    }
    \label{fig:ri3}
\end{figure}

\begin{figure}
    \includegraphics[width=1.0\columnwidth]{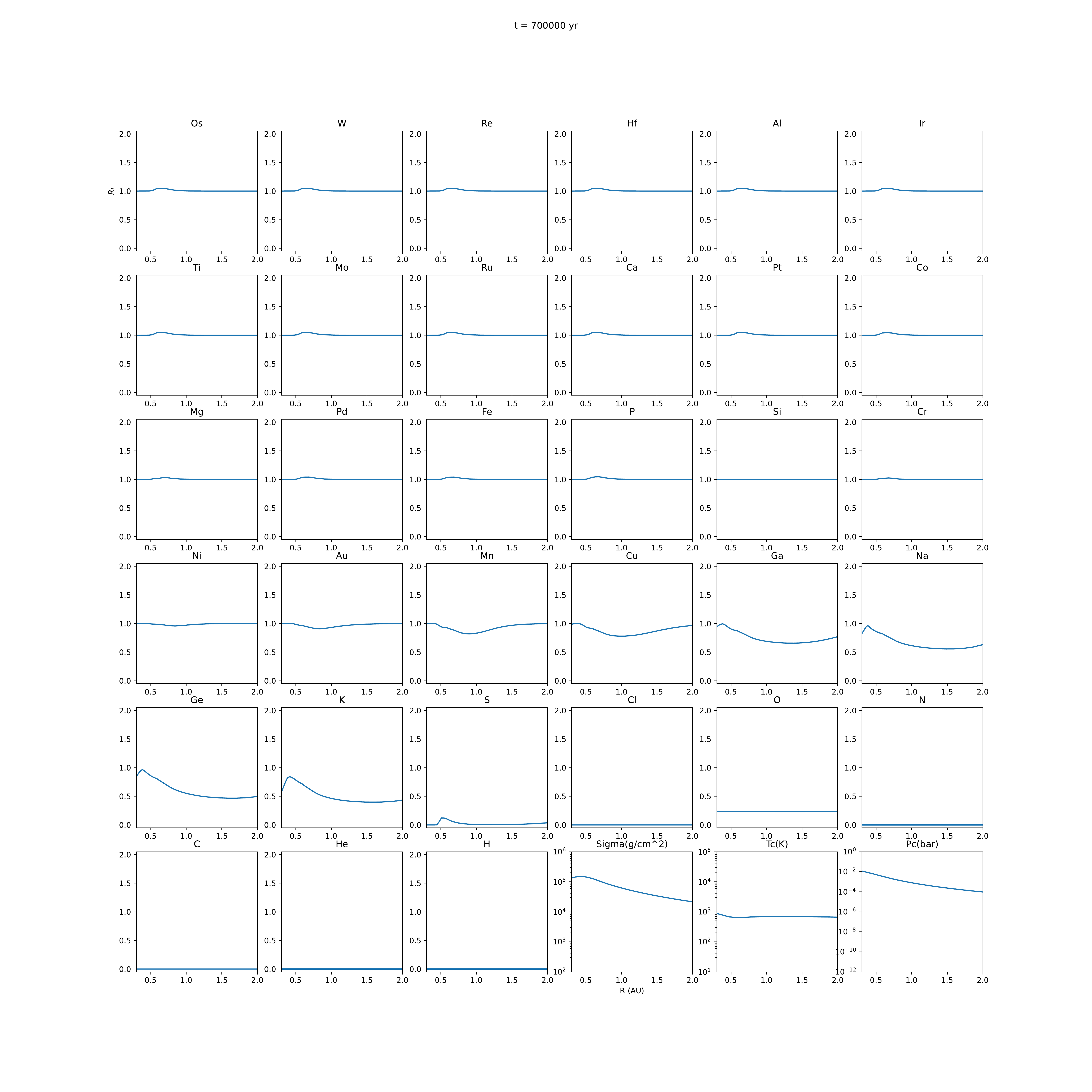}
    \caption{
    Snapshots for the abundance and disk properties at 0.7 Myr. The elements are ordered according to their T50s.
    }
    \label{fig:ri3new}
\end{figure}

\begin{figure}
    \includegraphics[width=1.0\columnwidth]{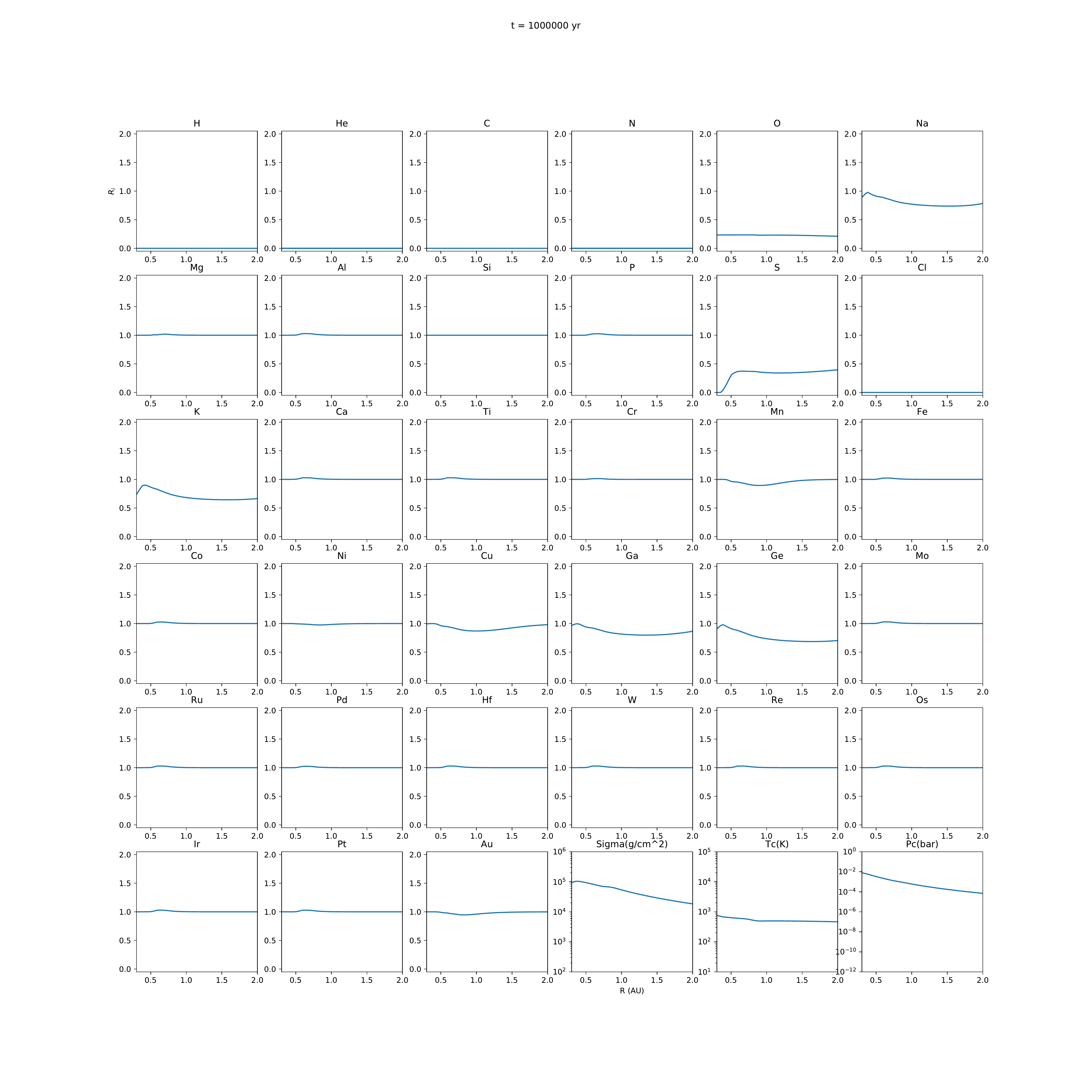}
    \caption{
    Snapshots for the abundance and disk properties at 1 Myr.
    }
    \label{fig:ri4}
\end{figure}

\begin{figure}
    \includegraphics[width=1.0\columnwidth]{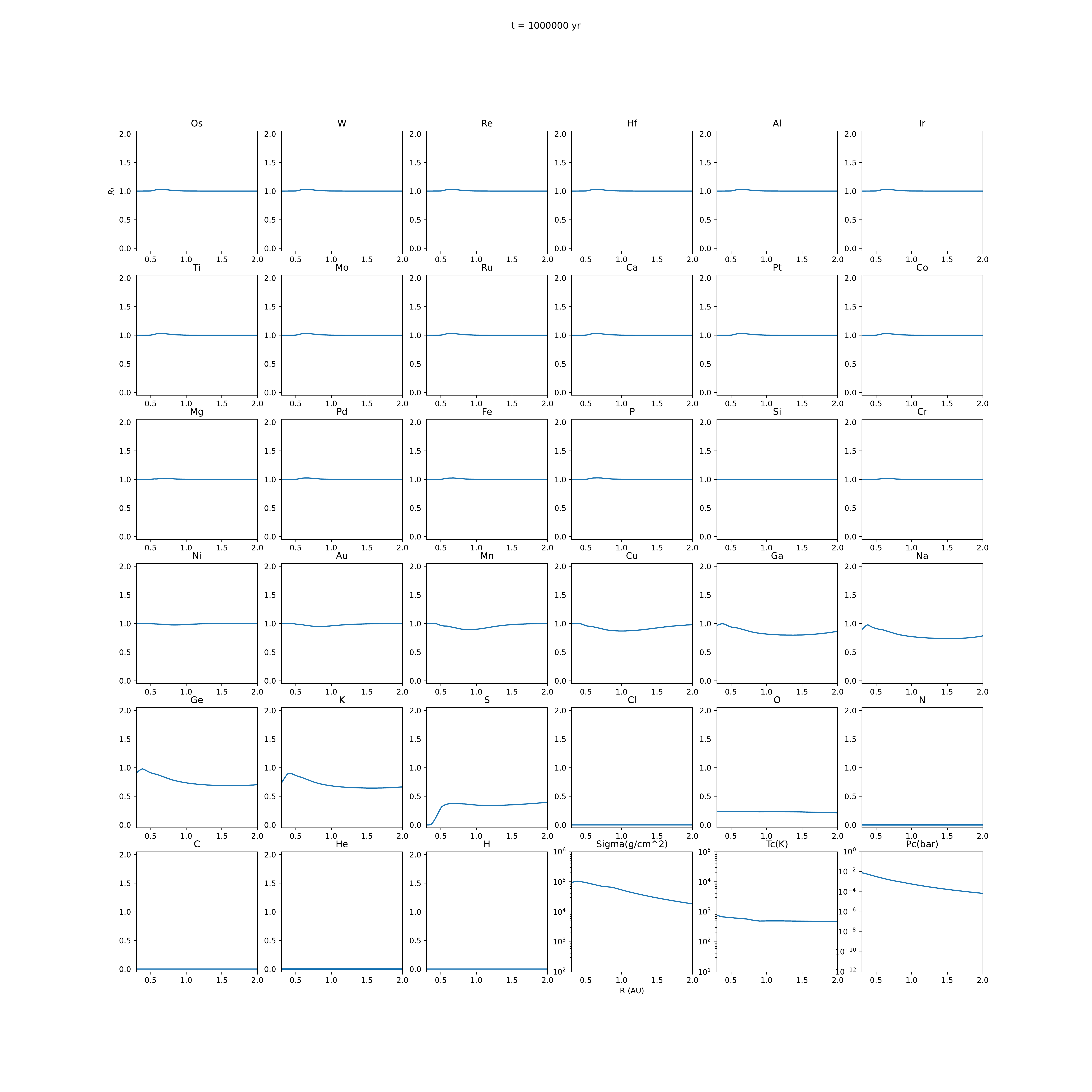}
    \caption{
    Snapshots for the abundance and disk properties at 1 Myr. The elements are ordered according to their T50s.
    }
    \label{fig:ri4new}
\end{figure}


\bibliographystyle{aasjournal}

\end{document}